\documentclass[twocolumn,superscriptaddress,floatfix,preprintnumbers, nofootinbib]
{revtex4-2} 
\pdfoutput=1
\usepackage[colorlinks=true,breaklinks=true]{hyperref}
\usepackage{float}
\usepackage{subcaption}
\usepackage[normalem]{ulem}
\usepackage[utf8]{inputenc}
\hypersetup{allcolors=[rgb]{0.0 0.0 0.6},linkcolor=[rgb]{0.75 0.05 0.05}}
\usepackage{amsmath,amssymb} \usepackage{epsfig}  
\usepackage{graphicx}   
\usepackage{color}
\usepackage{multirow}
\usepackage{comment}
\usepackage{amssymb}
\usepackage{orcidlink}
\usepackage{soul}
\definecolor{ForestGreen}{RGB}{34,139,34}

\begin{document}

\title{Probing supermassive black hole scalarization with Pulsar Timing Arrays}

\smallskip

\author{Clemente~Smarra \orcidlink{0000-0002-0817-2830}}
\email{csmarra@sissa.it}
\affiliation{SISSA International School for Advanced Studies, Via Bonomea 265, 34136, Trieste, Italy}
\affiliation{INFN,  Sezione di Trieste, Via Bonomea 265, 34136, Trieste, Italy}
\affiliation{IFPU — Institute for Fundamental Physics of the Universe, Via Beirut 2, 34014 Trieste, Italy}

\author{Lodovico~Capuano \orcidlink{0009-0001-0369-6635}}
\email{lcapuano@sissa.it}
\affiliation{SISSA International School for Advanced Studies, Via Bonomea 265, 34136, Trieste, Italy}
\affiliation{INFN,  Sezione di Trieste, Via Bonomea 265, 34136, Trieste, Italy}
\affiliation{IFPU — Institute for Fundamental Physics of the Universe, Via Beirut 2, 34014 Trieste, Italy}

\author{Adrien~Kuntz \orcidlink{0000-0002-4803-2998}}
\email{adrien.kuntz@tecnico.ulisboa.pt}
\affiliation{CENTRA, Departamento de F\'{\i}sica, Instituto Superior T\'ecnico -- IST, Universidade de Lisboa -- UL, Avenida Rovisco Pais 1, 1049-001 Lisboa, Portugal}

\begin{abstract}
Scalar-tensor theories with a scalar field coupled to the Gauss-Bonnet invariant can evade no-hair theorems and allow for non-trivial scalar profiles around black holes. This coupling is characterized by a length scale $\lambda$, which, in an effective field theory perspective, sets the threshold below which deviations from General Relativity become significant. LIGO/VIRGO constraints indicate $\lambda$ is small, implying supermassive black holes should not scalarize. However, recent work suggests that scalarization can occur within a narrow window of masses, allowing supermassive black holes to scalarize, while leaving LIGO/VIRGO sources unaffected. We explore the impact of this scenario on the stochastic gravitational wave background recently observed by Pulsar Timing Arrays. We find that scalarization can alter the characteristic strain produced by circularly inspiralling supermassive black hole binaries and that current data shows a marginal preference for a non-zero $\lambda$. However, similar signatures could arise from astrophysical effects such as orbital eccentricity or environmental interactions, emphasizing the need for improved modeling and longer observations to discriminate among the different scenarios.
\end{abstract}

\maketitle
\section{Introduction}
Despite the great success of General Relativity (GR) in describing gravitational phenomena~\cite{Will_2014,LIGOScientific:2021sio}, there are both theoretical and phenomenological motivations to extend it into a more general framework.
Among the several possible modifications of GR, scalar-tensor (ST) theories have been studied in greater detail,  due to their conceptual simplicity and relevant phenomenology. All ST theories are characterized by the inclusion of at least one fundamental scalar degree of freedom (DOF) in the gravitational sector. Theories within this class provide an appealing explanation to phenomena related to the universe dynamics like dark energy or inflation~\cite{Riess_1998,Planck:2018vyg,Brax:2017idh,Starobinsky:1980te,Guth:1980zm,Olive:1989nu}.

On the other hand, it has been shown that most of the interaction terms admitted in (ghost-free) ST gravity do not affect black hole (BH) solutions at the background level. In other words, BHs in most ST theories are exactly the same as in GR, meaning that they are fully determined by mass, spin and, in principle, electric charge, while the additional scalar DOF provides no further ``charge" to the BH.
This fact is related to the well-known no-hair conjecture and has been rigorously formalized in some specific cases through no-hair theorems~\cite{Bekenstein:1971hc, Bekenstein:1972ky,Graham:2014mda,Hui:2012qt,Capuano:2023yyh}. \\

Nevertheless, such theorems rely on a set of crucial assumptions that cannot always be taken for granted. In particular, it has been shown that a specific interaction term, among all those leading to second-order equations of motion (EOM), can violate all known no-hair theorems, allowing for BH solutions with a non-trivial scalar profile~\cite{Sotiriou:2013qea,Sotiriou:2014pfa}. This term is given by $\lambda^2 F(\varphi) \mathcal{G}$, where $\lambda$ is a coupling constant with dimension of a length, $\varphi$ is the scalar field, $F$ is an analytic function, and $\mathcal{G}$ is the Gauss-Bonnet (GB) invariant
\begin{equation}
    \mathcal{G}=R_{\mu\nu\rho\sigma}R^{\mu\nu\rho\sigma}-4 R_{\mu\nu}R^{\mu\nu}+R^2\,.
\end{equation}
The interesting properties of this model include BH solutions coupled to the scalar field via the so-called scalar charge~\cite{Sotiriou:2013qea,Sotiriou:2014pfa,Doneva:2017bvd,Doneva:2021tvn,Eichhorn:2023iab}, which has been used to constrain the coupling parameter $\lambda$ using Gravitational Waves (GWs) observations from the LIGO/Virgo collaboration~\cite{LIGOScientific:2021sio}. 

An interesting feature of this model is that the best constraints on $\lambda$ actually come from the smaller BHs from which we observe the GWs. This is tied to the dimensionful nature of the coupling: the smaller the BH scale, the better the constraint on $\lambda$. In fact, one can view interaction terms coming with a dimensionful coupling constant $\lambda$, such as the scalar-GB coupling, as irrelevant operators in an effective field theory (EFT) perspective~\cite{Donoghue:2022eay}. This implies that the contribution of this operator is suppressed at scales that are much larger than $\lambda$, whereas deviations from GR are expected to become increasingly significant at smaller scales. For this reason, it is usually assumed that supermassive BHs (SMBHs) are the same as in GR, even in presence of the scalar-GB coupling.\\

The EFT interpretation of ST theories can also be made more concrete, as it can be shown that some specific operators arise as the low-energy limit of quantum gravity theories. This is the case for some scalar-GB couplings, that emerge in the low-energy limit of string theories~\cite{Polchinski:1998rq}.
However, in principle, one can suppose that high-energy effects, at a given scale, lead BHs back to the GR configuration, providing a well-delimited range of scales in which deviations from GR can occur.
This happens, for instance, in a recently proposed model~\cite{Eichhorn:2023iab}, in which SMBHs can possess a scalar charge while lower mass BHs are the same as in GR.
It is therefore reasonable to envision departures from GR in the signals that we receive from SMBHs, which, when coupled in binary systems, are considered the most likely source of the recently observed stochastic GW background (SGWB).\\

Indeed, in recent years, Pulsar Timing Array (PTA) collaborations reported evidence for a temporally-correlated stochastic process leaving its imprint on the times of arrival (TOAs) of radio beams coming from a network of monitored millisecond pulsars in our Galaxy~\cite{EPTAIII, Agazie_2023_SGWB, Reardon_2023, Xu_2023, Miles_2024}.  Furthermore, the correlations observed among different pulsars were found to be consistent with the Hellings-Downs (HD) pattern~\cite{1983ApJ...265L..39H}, as expected for a SGWB.
Although the observed data can also be explained with Early Universe processes (see e.g. Refs.~\cite{EPTA_IV, Afzal_2023}), the most likely interpretation consists in the incoherent superposition of GWs emitted by a population of inspiralling SMBH binaries (SMBHBs).

Assuming circular orbits — which we restrict to in the following — the characteristic strain spectrum associated with GW-driven orbital decay is expected to follow a power-law scaling as $f^{-n}$, with the spectral index $n = 2/3$, where $f$ denotes the emitted GW frequency~\cite{Phinney:2001, Jaffe:2002rt}. Eccentric orbits as well as the presence of any additional non-GW energy loss channel - such as dynamical friction or, in our case, scalar radiation- would accordingly modify the spectral index $n$~\cite{Enoki_2007, Huerta_2015, Chen:2016kax, Chen_2017, Sesana_2015, Taylor_2017, Burke-Spolaor:2018bvk}.
On the other hand, the predicted angular correlation pattern is a property of the polarization of GWs at the Earth location, and will essentially coincide with the HD curve whenever - as we consider in the following - the scalar field is not coupled to matter.\\

In this article, we hence study the modification of the SGWB predicted by an assembly of circular SMBHBs evolving under the ST theory of gravity studied in Ref.~\cite{Eichhorn:2023iab}. 
In Sec.~\ref{sec:AH}, we summarize the framework of black hole scalarization and lay the foundations for the analysis developed throughout the paper.
In Sec.~\ref{sec:PTA}, we review how to construct the SGWB from an astrophysical population of SMBHBs, and we compare the resulting characteristic strain in GR with that predicted by the ST theory described in Sec.~\ref{sec:AH}. 
In Sec.~\ref{sec:results}, we analyze the EPTA DR2New dataset, comprising the most recent 10.3 yr of data of the EPTA collaboration, and the NANOGrav 15yr data, looking for signatures of non trivial SMBHBs scalarization.
Finally, in Sec.~\ref{sec:concl}, we summarize our findings and conclude with a brief discussion of their implications and possible directions for future work. Additional technical details and supplementary plots are provided in the Appendices.
In the following, we will work in natural units $c = 1$, while keeping expressed the gravitational coupling constant $G$, and using the mostly plus signature of the metric tensor $(-,+,+,+)$.

\section{BH scalarization}\label{sec:AH}

\subsection{Motivation}
In this section, we review some relevant results in the context of BH scalarization, in order to provide a more solid motivation for our investigation. 

We focus on a model in which the scalar field is only ruled by a canonical kinetic term and a coupling with the GB invariant, neglecting other possible couplings to curvature, or other kind of higher derivative interactions. We additionally assume that the coupling of scalar to matter is negligible, as suggested by the stringent bounds from the Solar System~\cite{Bertotti:2003rm, Will_2014} which apply to models with a scalar-GB coupling \footnote{Notice that Solar System bounds apply in this scenario, as the effective Compton wavelength of the scalar field (see Eq.\eqref{KG_eq} later on) is larger than the typical distances probed by Solar System experiments. }. This means that specifying a Jordan or Einstein frame is unnecessary, as the metric is the same in both frames in the absence of matter coupling.
We will refer to this class of models as Einstein-scalar-Gauss-Bonnet (EsGB) gravity. The action of EsGB can be expressed in the simple form 
\begin{equation}
    S_{\rm EsGB} =\frac{1}{16\pi G}\int{\rm d}^4x \sqrt{-g}\,\left[R- 2 \left(\partial \varphi \right)^2+\lambda^2F(\varphi)f\left(\mathcal{G}\right)\right]\,,
    \label{EsGB_Action}
\end{equation}
where $\lambda$ and $F$ are the dimensionful coupling and the function introduced in the previous section, and $f$ is an analytic function of the GB invariant.\\

Let us first focus on the case in which $f(\mathcal{G})=-\mathcal{G}$. An important distinction must be made among models within this class, depending on the form of $F(\varphi)$. In fact, if the first derivative $F'(\varphi_0)\neq 0$ for every value $\varphi_0$ of the scalar field, one obtains a theory in which all possible BH solutions source a nontrivial scalar profile~\cite{Sotiriou:2013qea,Sotiriou:2014pfa}. This means that GR BH solutions are not solutions of EsGB. This is the case, for instance, for the shift-symmetric \cite{Sotiriou:2013qea,Sotiriou:2014pfa} and dilatonic \cite{Kanti:1995vq,Pani:2009wy,Kleihaus:2015aje} coupling functions, given by $F(\varphi)=\varphi$ and $F(\varphi)=\exp{(\varphi)}$, respectively.

Notice that the shift-symmetric models exhibit characteristic features, like finite-radius singularities, whose location depends on the coupling $\lambda$~\cite{Sotiriou:2013qea,Sotiriou:2014pfa}. The presence of these singularities, if one requires them to be always located within the event-horizon, imposes a lower bound on the BH size, that translates into an observational constraint on the coupling $\lambda$ for this theory~\cite{Charmousis:2021npl,Fernandes:2022kvg}. However, it is worth mentioning that the shift-symmetric EsGB theory, when requiring that it produces observable deviations from GR at astrophysical scales, presents issues with causality~\cite{Serra:2022pzl}.
In the dilatonic case, one also encounters a lower BH mass limit, set by the dimensionful coupling $\lambda$, so that the best experimental constraints can be put from low mass BH observations \cite{Maselli:2014fca}.

A different situation occurs if one imposes $F'(\varphi_0)= 0$ for a certain $\varphi_0$~\cite{Silva:2017uqg}. In this case, BH solutions typically exist in separate branches. In particular, the GR BH solutions, characterized by a trivial configuration of the scalar field, given by $\varphi = \varphi_0$, are also allowed. 
Consider now the dynamics of a scalar perturbation $\delta\varphi$ around this configuration, ruled by the linearized Klein-Gordon equation 
\begin{equation}
    \left(\Box-\frac{\lambda^2\mathcal{G}}{4}F''(\varphi_0)\right)\delta\varphi=0\,,
    \label{KG_eq}
\end{equation}
where both the D'Alembertian operator $\Box\equiv g^{\mu\nu}\nabla_\mu\nabla_\nu$ and the GB invariant $\mathcal{G}$ are computed on the GR BH solution.

As it is clear from Eq.~\eqref{KG_eq}, the coupling with the GB invariant acts as an effective mass for the scalar perturbation, and, therefore, its sign determines the stability of the GR configuration. In particular, a tachyonic instability is developed if the following condition is verified:
\begin{equation}
    F''(\varphi_0)\mathcal{G}<0\,.
\end{equation}
If this is the case, GR BHs naturally evolve towards other branches of solutions, in which the scalar field exhibits a nontrivial configuration. This phase transition is called BH scalarization~\cite{Silva:2017uqg,Doneva:2017bvd,Doneva:2021tvn}.  
Some instances of models allowing for scalarization are given by $\mathcal{Z}_2$-symmetric polynomial couplings~\cite{Silva:2017uqg,Julie:2023ncq}, or the exponential model~\cite{Doneva:2017bvd,Doneva:2021tvn} given by
\begin{equation}
    F(\varphi)=\frac{1}{2\beta}\left[1-\exp{(-\beta \varphi^2)}\right]\,.
\label{exp_model}
\end{equation}
Remarkably, the latter does not predict any lower bound to BH masses.\footnote{In the present discussion, we only consider solutions that do not exhibit any node in the radial profile of the scalar field. However, one can find several branches of solutions with $n\geq1 $ nodes, which instead exist in a finite mass range.} Moreover, for these models, unlike the shift-symmetric case, there is no clear causality argument restricting the scale at which deviations from GR can be expected \cite{Serra:2022pzl}.
Nevertheless, BH solutions with a mass smaller than a critical value have been shown to be affected by hyperbolicity loss~\cite{Blazquez-Salcedo:2022omw}. This phenomenon signals the breakdown of the regime of validity of the theory. For this reason, the exponential model of BH scalarization should be considered only for BH masses that are larger than this threshold. On the other hand, in an EFT perspective, moving towards a higher energy regime (or, equivalently, smaller length scale), one should complement the action with additional terms, which can in principle affect BH solution and prevent scalarization of small BHs.\\

A practical example of how higher-order terms in the action can dramatically affect the mass range of scalarized BH solutions is presented in Ref.~\cite{Eichhorn:2023iab}, where the exponential form of Eq.~\eqref{exp_model} is considered for $F(\varphi)$, and the coupling function appearing in Eq.~\eqref{EsGB_Action} is replaced by
\begin{equation}
   f(\mathcal G) = - \mathcal G + \tilde \lambda^4 \mathcal G^2\,,
\label{Aaron_model}
\end{equation}
with $\tilde{\lambda}$ being another dimensionful coupling constant. In this model, to ensure that the Schwarzschild BH still solve the EOMs, one has to restrict to the case $\varphi = 0$. In the $\tilde\lambda =0$ case, the Schwarzschild solution is admitted with any constant $\varphi_0$, as $\mathcal{G}$ yields a trivial contribution to the EOM, being it a surface term. In the more general case $\tilde\lambda\neq0$, the condition for the tachyonic instability to occur is now given by
\begin{equation}
    f(\mathcal G)F''(0)<0\,.
\end{equation}
Hence, due to the presence of two different powers of the GB invariant in Eq.~\eqref{Aaron_model}, this model allows for BH scalarization in a finite window of BH masses. 

This practically demonstrates that higher-order terms in EsGB can, in principle, protect BHs from scalarization at small scales, where the standard exponential model described in Ref.~\cite{Doneva:2017bvd} is affected by hyperbolicity loss, while allowing for this phenomenon to occur at larger scales. For this reason, the constraints on scalar charges provided by LIGO-Virgo do not necessarily imply that scalarization should be excluded for SMBHs. \\

In the following, we will be agnostic about the specific model,~\footnote{Notice that, in a work released after our initial submission, the model presented in Ref.~\cite{Eichhorn:2023iab} has been argued to be inconsistent from an EFT point of view~\cite{Thaalba:2025ljh}.  We stress that our reasoning remains valid regardless of the specific model, as long as scalarization takes place at supermassive scales.} and just assume that the scalar field is coupled to some positive powers of the GB invariant. We will also assume that scalarization happens in a finite window of BH masses, centered at supermassive scales ($\sim 10^9\,M_{\odot}$), in light of the discussion above.

Additionally, we assume that neutron stars are not scalarized. As discussed in Ref.~\cite{Eichhorn:2023iab}, this is likely to be the case if the dimensionful couplings in the theory lie at supermassive scales. We point out that, in principle, having scalarized pulsars could result in modifications of their orbits in binary systems. However, such effect should be accounted for within the pulsar timing model used in the PTA pipeline.
A detailed treatment of this aspect for the theory under examination is beyond the scope of the present work and is left for future study.

We finally note, that in an EFT perspective, one should also consider couplings between the scalar field and other curvature terms, like the Ricci scalar $R$. In particular, the inclusion of a coupling in the form $H(\varphi)R$ in the standard EsGB theory, has a non-trivial impact on scalarization, and allows to partially tame the loss of hyperbolicity in some regimes~\cite{Antoniou:2021zoy,thaalba2024dynamicssphericallysymmetricblack,PhysRevD.109.L041503,PhysRevD.111.024053,Antoniou_2022}. We will, however, not consider this kind of term in the present discussion, leaving it for future investigation. 

\subsection{Scalar charge model}\label{sec:scacha}
Given the motivation provided in the previous section, we now want to test the presence of a nontrivial scalar field profile around SMBHs. In particular, we want to assess the possible impact of such deviations from GR on GWs emitted by SMBHB inspirals.

In general, an object sourcing a nontrivial scalar profile in a ST theory is characterized at large distance $r$ by a scalar charge. More quantitatively, we define the dimensionless scalar charge $\alpha$ as the coefficient of the $GM/r$ term in an expansion of the scalar field $\varphi$ at infinity, where $M$ stands for the mass of the BH under examination. Namely:
\begin{equation}
    \varphi = \varphi_{\infty}+\frac{ \alpha G M}{r}+\mathcal{O}\left(\frac{1}{r^2}\right)\,,
\end{equation}
where $\varphi_{\infty}$ is the background value of the scalar field, which we take to be vanishing.

As we will explain in the next section, the scalar charge affects the time evolution of the frequency of the GWs emitted by a SMBHB system. The effect is twofold: on the one hand, the scalar charge directly affects the power emitted through gravitational radiation (dominated by the quadrupole term); on the other hand, if the scalar charges of the two BHs in the system are different, the binary also emits power through scalar radiation, which is dominated by a dipole term. 
\begin{figure}[t]
    \centering
    \includegraphics[width = 0.45\textwidth]{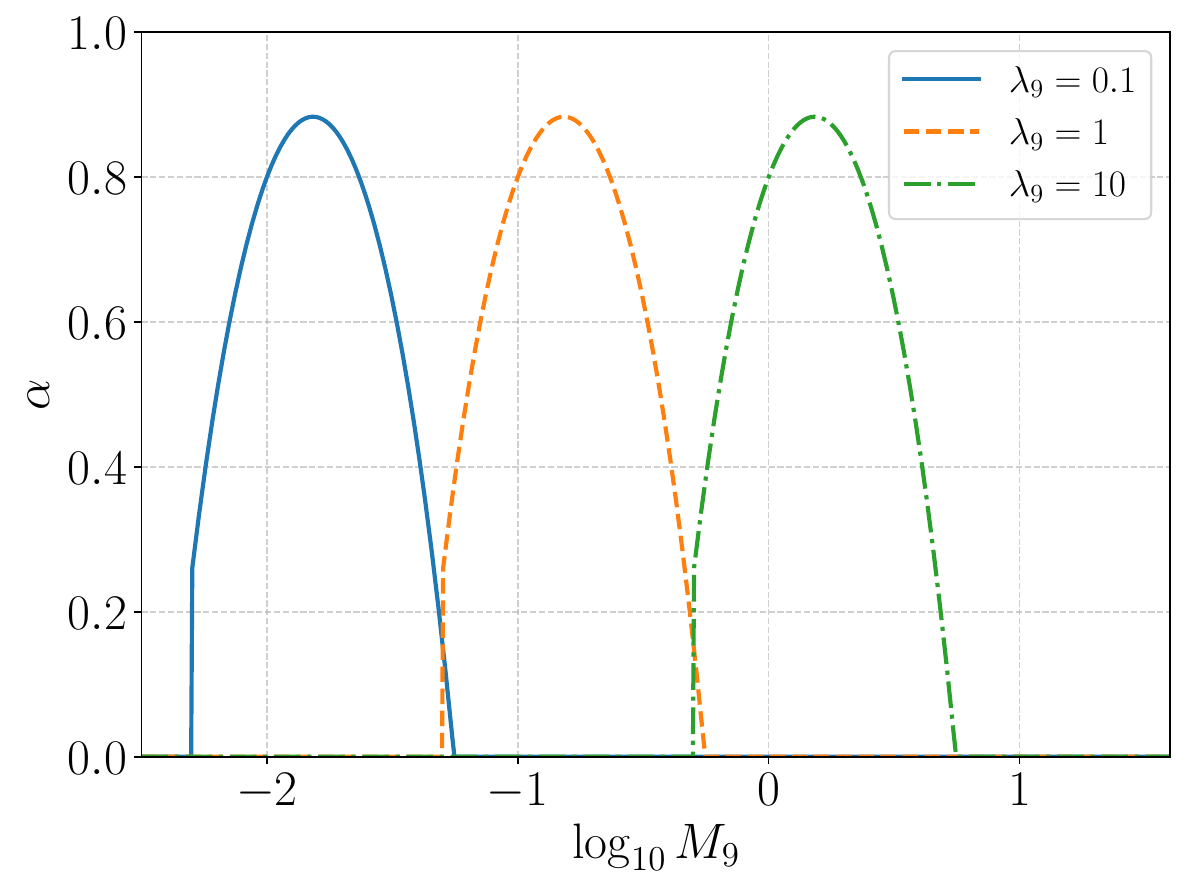}
    \caption{Phenomenological model for the dimensionless scalar charge, given by Eq.~\eqref{eq:PhenoCurve}.  We consider different values of the parameter $\lambda_9$, corresponding to models with BH scalarization at different mass scales. The lower mass cutoff is motivated by the presence of a minimum mass in the model of Ref.~\cite{Eichhorn:2023iab}, and by the fact that in standard EsGB we have hyperbolicity loss below a given mass value. }
    \label{fig:b5}
\end{figure}\\

In order to provide a description of the signal emitted by a population of SMBHBs, we need to model the dependence of the scalar charge on the BH mass. This procedure is possible as in EsGB we are dealing with \textit{secondary} BH hairs. In other words, the scalar charge is not an independent parameter, but rather a function of mass $M$ and scalar field at infinity $\varphi_\infty$. Of course, this dependence is strongly model-specific, is likely to be affected by the BH spin and, in general, has no closed analytic form. We then follow a phenomenological approach, motivated by the discussion in the previous section. 

In more detail, we model the scalar charges with the functional form
\begin{equation}
    \alpha(x)=\text{max}\{0,\tilde{\alpha}(x)\}\,,
    \label{eq:PhenoCurve}
\end{equation}
with
\begin{equation}\label{eq:tildealpha}
    \tilde{\alpha}(x)\equiv\frac{\lambda_9}{M_9}\left(0.2-3\left(\frac{M_9}{\lambda_9}-0.3\right)^2\right)\Theta_\text{H}\left(\frac{M_9}{\lambda_9}-0.05\right)\,,
\end{equation}
where $\Theta_\text{H}$ is the Heaviside step-function and we defined for later convenience $M_9 \equiv M/(10^9M_{\odot})$ 
and $\lambda_9 \equiv \lambda/(10^9M_{\odot})$. This curve is represented in Fig.~\ref{fig:b5}. Notice that the model displays an upper mass bound at the point at which the function $\tilde{\alpha}$ becomes negative. As far as we now, such upper mass bound is present in all scalarization models in the literature. On the other hand, the Heaviside function produces a sharper cutoff at lower masses.

Our choice of the functional form in Eq.~\eqref{eq:tildealpha} is intended as a minimal model that captures the key features of scalarization at the SMBH scale discussed so far: namely, the existence of a mass window bounded both from below and above, along with a maximum scalar charge at intermediate scales. We emphasize that the adopted parametrization is not meant to reproduce the precise dynamics of any specific ST theory, but rather to serve as a flexible template for studying potential observational imprints in the the PTA data, as we will do below.

Moreover, in our analysis, we adopt the value of the cutoff in Eq.~\eqref{eq:tildealpha} as a proxy for the model-dependent scalarization threshold identified in Ref.~\cite{Eichhorn:2023iab}.
This choice is intentionally made to emphasize the potential impact of scalarization on the SGWB by allowing for a wider scalarization window, while preserving the qualitative features of the underlying theory. Choosing higher cutoff values, consistent with Fig.~2 of Ref.~\cite{Eichhorn:2023iab}, would yield a narrower window and a correspondingly weaker imprint on the PTA signal.
To highlight the observational signatures, we focus on cutoff values that yield detectable effects in current PTA data. 
Future observations, with longer time spans and improved precision, will be able to test more restrictive thresholds.
Finally, notice that care should be taken when comparing Fig.~\ref{fig:b5} to the results in Ref.~\cite{Eichhorn:2023iab}, as we are representing the dimensionless scalar charge, rather then the dimensionful scalar charge divided by $\lambda$.

\section{Spectral shape of the SGWB in EsGB}\label{sec:PTA}
PTAs are facilities timing an ensemble of Galactic pulsars for decades, with an approximate cadence of $\sim 1 \,\text{week}$. The inverse of the observational timespan sets the minimum frequency PTAs are sensitive to, roughly $1\, \text{nHz}$, while the inverse cadence sets the maximum frequency, around $1 \,\mu\text{Hz}$. 

The TOAs of the radio beams coming from pulsars are then fitted with a \textit{timing model}, which includes several variables, describing the pulsar spin period, the spin period derivative, its sky localization, the effect of the interstellar medium, the impact of bodies in the Solar System etc~\cite{Hobbs:2006, Edwards:2006, Lentati:2013, Lorimer_2004}. The difference between the observed TOAs and the ones predicted by the timing model fit is termed \textit{timing residuals}, and will be labeled as $\delta t$ in the following. The \textit{timing residuals} contain unmodeled information: aside from measurement noise, they can also include the imprints of new physical phenomena.

Recently, PTA collaborations have reported evidence for the presence of an underlying temporally-correlated common stochastic process in the timing residuals of pulsars~\cite{EPTAIII, Agazie_2023_SGWB, Reardon_2023, Xu_2023, Miles_2024}, with a distinctive pattern of inter-pulsar correlation resembling the HD curve~\cite{1983ApJ...265L..39H}. These features point to the presence of a SGWB, inducing variations in the TOAs of radio beams from the pulsar ensemble. Strong efforts have been dedicated to unveil the origin of such a background. While it could be produced by cosmological processes, such as cosmic strings or phase transitions in the Early Universe~\cite{EPTA_IV, Afzal_2023}, the most likely explanation of the SGWB lies in the incoherent superposition of GWs emitted by a population of inspiralling SMBHBs at low redshift ($z \lesssim 2$). \\

In this work, we assume indeed such an \textit{astrophysical} origin of the observed signal.
In particular, we restrict to an isotropic, unpolarized, stationary and gaussian background, which is fully characterized by the two point function of timing residuals~\cite{Maggiore:2018sht}:
\begin{equation}
    \left<\delta t_a(t) \delta t_b(t)\right> \sim \xi(\theta_{ab})\int_{f_l}^{f_h}{\rm d}f\,P(f),
    \label{eq:dtdt}
\end{equation}
where the indices $a$ and $b$ label the pair of pulsars being correlated, $\theta_{ab}$ the angle between them and $f_l$ and $f_h$ the minimum and maximum frequency detectable by PTAs, respectively. The angular dependence of the two point function is completely factored out of the integral in the HD curve, $\xi (\theta_{ab})$, defined as:
\begin{equation} \label{eq:angularcorr}
    \xi(\theta_{ab}) \equiv \frac{3}{2}r_{ab} \log r_{ab} -\frac{1}{4} r_{ab} + \frac{1}{2},
\end{equation}
with $r_{ab} = (1 - \cos\theta_{ab})/2$. The spectral features of the correlation are instead described by the power spectrum of timing residuals, $P(f)$, which can be expressed as~\cite{Maggiore:2007ulw}
\begin{equation}
    P(f) = \frac{h_c^2(f)}{12\pi^2 f^3},
    \label{eq:pstim}
\end{equation}
where the quantity $h_c(f)$ is commonly referred to as the \textit{characteristic strain} or \textit{characteristic GW amplitude} of the SGWB. The advantage of the formulation in Eq.~\eqref{eq:pstim} lies in the fact that $h_c(f)$ is an adimensional quantity. For a population of inspiralling SMBHBs,  the characteristic strain can be expressed as \footnote{The derivation of the characteristic strain is the same as the one performed in GR~\cite{Phinney:2001}, since, as it is shown in \cite{Stein:2010pn}, modified gravity theories including a coupling of a dynamical scalar field to high-rank curvature invariants produce the same effective stress-energy tensor of GWs as the GR one.} :
\begin{equation}
    \begin{split}
	h_c^2(f) = &\frac{4G}{\pi f^2}\int_0^\infty \frac{{\rm d}z}{1 + z} \times\\
    \times&\int {\rm d}\chi ~\frac{{\rm d}n}{{\rm d}z{\rm d}\chi}(z; \chi) \left. \left(\frac{{\rm d}E_\text{gw}^s(f_s; \chi)}{{\rm d}\text{log} f_s}\right)\right|_{f_s = (1 + z ) f}.
    \end{split}
	\label{eq:hc}
\end{equation}
Let us now explain the ingredients forming Eq.~\eqref{eq:hc}:
\begin{itemize}
    \item $\frac{{\rm d}n}{{\rm d}z{\rm d}\chi}(z; \chi)$: this is the \textit{comoving merger density} of sources at redshift $z$, characterized by a specific set of parameters $\chi$. In particular, $\chi$ can label the chirp mass of the system, the mass ratio, the spins or whatever source parameter is relevant for the analysis;
    \item $\frac{{\rm d}E_\text{gw}^s(f_s; \chi)}{{\rm d}\text{log} f_s}$: this quantity is the source-frame GW spectrum emitted by SMBHBs at redshift $z$, with source parameters $\chi$. The GW emission frequency $f_s$ is then measured by PTAs at the redshifted frequency $f = f_s/(1 + z)$;
    \item $(1 + z)^{-1}$: redshift of the emitted energy $E_\text{gw}^s(f_s; \chi)$.
\end{itemize}
The formula in Eq.~\eqref{eq:hc} clearly shows that the frequency dependence of the characteristic strain entirely comes from the spectrum of the emitted energy in GW, while the merger density contributes to setting its normalization. 
In particular, dipolar scalar emission in EsGB gravity modifies the GW spectrum and thus the frequency dependence of the strain, as we will show later in this section. On the other hand, the angular correlation pattern in Eq.~\eqref{eq:angularcorr} is unaffected, since we neglect the coupling of the scalar to matter in this work. Indeed, in this case, the Jordan and Einstein frame metrics coincide. This means that matter follows the geodesics of a metric which propagates only two polarizations in vacuum, as it can easily be checked that the action~\eqref{EsGB_Action} does not mix scalar and tensor modes at quadratic order in perturbations. 
As a result, we can adopt standard GR formulae, from which eq.~\eqref{eq:angularcorr} follows.

Let us now analyze the formula~\eqref{eq:hc} in greater detail. First, we describe an analytical model of the comoving merger density; then, we focus on the frequency behavior of $h_c(f)$ in GR and EsGB gravity.

\subsection{Comoving merger density}\label{sec:com}

In order to compute the characteristic strain in Eq.~\eqref{eq:hc}, we need to specify the comoving merger density, which informs us about the underlying astrophysical population of SMBHBs. This quantity is among the main outputs of semi-analytic models, see e.g.~\cite{Wyithe_2003, Berti_2008, Sesana_2008, Barausse_2012, McWilliams_2014}. 

In this work, we adapt the analytic merger density prescription developed in Ref.~\cite{Middleton_2015}, which successfully captures the key features of semi-analytic models, while remaining entirely agnostic about the framework-specific merger history of SMBHBs. This allows us to focus exclusively on the aspects directly constrained by PTA data, and not those that require assumptions about a specific merger history. 
Notice that semi-analytic models usually track the evolution of dark matter merger trees, and then empirically relate the merger density of galactic halos to the merger density of SMBHBs. Thus, as long as the scalar coupling to matter is negligible, the astrophysical model developed in Ref.~\cite{Middleton_2015} is totally insensitive to the underlying theory of gravity.

The merger density is hence given by
\begin{equation}
    \begin{split}
    \frac{{\rm d}n}{{\rm d}z{\rm d}\chi} = &\dot{n}_0\frac{{\rm d}t_s}{{\rm d}z}\left[\left(\frac{\mathcal{M}}{10^7 \text{M}_\odot}\right)^{-\gamma} \exp\left(-\mathcal{M}/\mathcal{M}_\star\right)\right] \times \\
    &\times \left[\left(1 + z\right)^\beta \exp\left(-z/z_0\right)\right]p(\log_{10}q),
    \end{split}
    \label{eq:middl}
\end{equation}
which interpolates among the results of different semi-analytic codes. We now describe the meaning of the parameters entering in this equation. In particular, $\mathcal M$ labels the chirp mass of the SMBHB system, $q$ its mass ratio, $d\chi = d\log_{10}\mathcal{M}\,d \log_{10}q$ and
\begin{equation}
    \frac{{\rm d}t_s}{{\rm d}z} = \frac{1}{H_0(1 + z) \left(\Omega_{m,0}(1 + z)^3 + \Omega_{\Lambda,0}\right)^{1/2}},
\end{equation}
with $H_0$ the Hubble constant, $\Omega_{m,0}$ and $\Omega_{\Lambda,0}$ the matter and dark energy densities at present times. 
As a benchmark, we take $H_0 = 70\, \text{km s}^{-1}\text{Mpc}^{-1}$, $\Omega_{m, 0} = 0.3$ and $\Omega_{\Lambda, 0} = 0.7$. We can factor Eq.~\eqref{eq:middl} in an overall amplitude, with $\text{Mpc}^{-3}$ dimensions, a redshift-dependent term and a mass-dependent term. Moreover, we consider a normalized distribution $p(\log_{10}q)$, to account for the fact that different systems may have different mass ratios $q$. In particular, we take it to be a log-uniform distribution in the range $\log_{10} q \in [-1.5, 0]$, which is broadly in agreement with previous works \cite{Kelley_2016, satopolito2023, satopolito2025}. As we will show later in Fig.~\ref{fig:EPTADR2New_9}, using a uniform distribution in the same mass ratio range, namely $q\in [0.03, 1]$, does not appreciably impact the results. However, notice that if the mass ratio distribution is strongly peaked around $q = 1$, the effect of EsGB gravity will be significantly suppressed. In this limit, for nearly equal-mass binaries, the difference in the scalar charges of the two components becomes negligible, leading to a strong suppression of the dipolar scalar emission (see Eq.~\eqref{app:etot} below). In the following, we will proceed under the \textit{caveat} that the true astrophysical distribution does not strongly favor equal mass binaries. Therefore, unless otherwise specified, we will just assume $\log_{10} q \in [-1.5, 0]$, as described before. \\

We now sum up the priors on the parameters of Eq.~\eqref{eq:middl}, with a short description of each of them:
\begin{itemize}
    \item $\log_{10} \frac{\dot{n}_0}{\text{Mpc}^{-3}\text{Gyr}^{-1}}  \in [-12, 7]$:  $\dot{n}_0$ is the merger density rate, and sets the overall normalization of the comoving merger density. %It is expressed in units of $\text{Mpc}^{-3}\text{Gyr}^{-1}$.
    \item $\beta \in [-2,7]$: this parameter controls the low-redshift power-law slope of the merger density.
    \item $z_0 \in [0.2, 5]$: this parameter describes the high-redshift behavior of the merger density. Together with $\beta$, it allows for the redshift evolution of the merger density.
    \item $\gamma \in [-3,3]$: it accounts for the distribution in chirp mass of the merger density.
    \item $\log_{10}\mathcal{M}_\star/\text{M}_\odot \in [6, 10]$: it gives an exponential cutoff in chirp mass.
\end{itemize}
Priors are mainly taken from Ref.~\cite{Middleton_2015}, except for the ones on $\log_{10} \frac{\dot{n}_0}{\text{Mpc}^{-3}\text{Gyr}^{-1}}$ and $\log_{10}\mathcal{M}_\star/\text{M}_\odot \in [6, 10]$, which are slightly modified to take into account recent measurements~\cite{EPTA_IV}.

\subsection{Characteristic strain in GR and EsGB gravity}
The second ingredient entering  Eq.~\eqref{eq:hc} is the GW energy spectrum, ${\rm d}E_\text{gw}^s/{\rm d}\text{log} f_s$.  At this stage, the fact that SMBHs acquire a scalar charge in EsGB theory becomes relevant.
Let us consider a system of two BHs, with masses $M_A$ , $M_B$ and dimensionless scalar charges $\alpha_A$, $\alpha_B$. We label the total mass as $M = M_A+M_B$.
In full generality, the GW spectrum can be written as 
\begin{equation} \label{eq:dEdf}
    \frac{{\rm d}E_\text{gw}^s}{{\rm d}\text{log} f_s} = \frac{{\rm d}E_\text{gw}^s}{{\rm d}t_s} \frac{{\rm d}t_s}{{\rm d}\log f_s},
\end{equation}
where we omitted the dependence on the redshift and source parameters for simplicity.
The quantity ${\rm d}E_\text{gw}^s/{\rm d}t_s$ is the \textit{instantaneous GW power} emitted in GWs in the source rest-frame.  From Appendix~\ref{app:binary_dynamics}, it is given in EsGB gravity at lowest post-Newtonian (PN) order by
\begin{equation}
	\frac{{\rm d}E_\text{gw}^s(f_s)}{{\rm d}t_s} = \frac{32}{5G_\mathrm{eff} (1+\alpha_A \alpha_B)} x^5,
	\label{eq:power}
\end{equation}
where we introduced the effective gravitational coupling  $G_{{\rm eff}}=G(1+\alpha_A\alpha_B)$, the chirp mass $\mathcal{M}\equiv(M_A M_B)^{3/5}/M^{1/5}$, and the parameter $x\equiv\left(\pi f_s G_\mathrm{eff} \mathcal{M} \right)^{2/3}$.

The above expression closely resembles the GR formula, with just an overall factor depending on $1+\alpha_A \alpha_B$, renormalizing the flux of energy in GWs.
However, the other ingredient entering in Eq.~\eqref{eq:dEdf}, namely the \textit{residence time} in a log-frequency bin $dt_s/d \mathrm{log} f_s$, differs significantly between GR and EsGB gravity. 

To begin, let us consider the case of pure GR, where the scalar charges are $\alpha_A=\alpha_B=0$.
By equating the GW energy flux in Eq.~\eqref{eq:power} to the time-derivative of the Keplerian binding energy of the two SMBHs,  the residence time in a log-frequency bin can be determined as: 
\begin{equation}
      \frac{{\rm d}t_s}{ {\rm d}\log f_s} = \frac{5}{96 x^4} G \mathcal{M}.
      \label{eq:res}
\end{equation}
Multiplying Eq.~\eqref{eq:power} and Eq.~\eqref{eq:res} and putting it back into Eq.~\eqref{eq:hc} yields, in GR,
\begin{align}
    &h_c^2(f) = A_\text{GR} f^{-4/3}, \nonumber\\
     &A_\text{GR} = \frac{4}{3\pi^{1/3}}\int_0^\infty \frac{{\rm d}z}{\left(1 + z\right)^{1/3}} \int {\rm d}\chi ~\frac{{\rm d}n}{{\rm d}z{\rm d}\chi} (z; \chi)\left(G\mathcal{M}\right)^{5/3}.
    \label{eq:grhc}
\end{align}
As specified below Eq.~\eqref{eq:hc}, the underlying astrophysical model plays a role in setting the normalization of $h_c$, while the theory of gravity (GR in this case) defines the frequency scaling, characterized by a spectral index $n = 2/3 $. 

Let us now consider EsGB gravity. In this case, the frequency evolution is modified due to an additional energy loss of the binary in the scalar sector, so that (see Appendix~\ref{app:binary_dynamics}):
\begin{align} \label{eq:gbf}
    \frac{\mathrm{d}\log f_s  }{\mathrm{d}t_s} &=  \frac{x^3}{ G_\mathrm{eff}\mathcal{M} (1+\alpha_A \alpha_B)} \bigg[  \frac{q^{2/5}}{(1+q)^{4/5}} \Delta\alpha^2 \nonumber \\
    &+ \frac{4x}{5} \big(24+(\alpha_A+\alpha_B)^2 \big) \bigg] \; ,
\end{align}
where where $q$ is the \textit{mass ratio}, defined as $q = M_B/M_A$, with $M_A$ the largest mass. and $\Delta\alpha = \alpha_A-\alpha_B$ is the difference between the two dimensionless scalar charges. The first term in this formula corresponds to a dipolar scalar emission, while the second term is a quadrupolar emission in both the scalar and the gravitational sector.  

\subsection{Qualitative analysis of the spectrum}\label{sec:qual}

We now want to discuss the main qualitative features of the SGWB in EsGB gravity, simplifying the physics as much as possible.
To do so, let us assume that all the SMBHBs in the PTA band, or at least the loudest ones building the bulk of the signal, are in a dipolar-emission-dominated regime. Which is the implication for the SGWB? To answer this question, we assume that, in this limit, the residence time in a log-frequency bin is entirely given by the first term in Eq.~\eqref{eq:gbf}, from which we find:
\begin{equation}
\begin{split}
    \frac{{\rm d}t_s}{{\rm d}\log f_s} &= \frac{ G_\mathrm{eff}\mathcal{M}(1+\alpha_A \alpha_B)\left( 1 + q\right)^{4/5}}{x^{3}q^{2/5}\Delta\alpha^2} \\
    &=\frac{\left( 1 + q\right)^{4/5}}{q^{2/5}G\mathcal{M}}\frac{1}{\Delta\alpha^2\pi^2 f_s^2},
    \end{split}
\end{equation}
Combining it with Eq.~\eqref{eq:power} and inserting the product in Eq.~\eqref{eq:hc} yields
\begin{equation}
    \begin{split}
    h_c^2(f) = &A_\text{GB} f^{-2/3},\\
    A_\text{GB} = &\frac{128\pi^{1/3}}{5}\int_0^\infty {\rm d}z\,\left(1 + z\right)^{1/3} \int {\rm d}\chi ~\frac{{\rm d}n}{{\rm d}z{\rm d}\chi}(z; \chi) \times\\
    \times & \left(G\mathcal{M}\right)^{7/3} \frac{\left(1 + q\right)^{4/5}}{q^{2/5}\Delta\alpha^2} (1 + \alpha_A \alpha_B)^{4/3}.
    \end{split}
    \label{eq:gbhc}
\end{equation}
As we can immediately see, the frequency slope of the characteristic strain is different than the one in GR (see Eq.~\eqref{eq:grhc}), providing us  with a distinct prediction of the EsGB theory. However, as Eq.~\eqref{eq:gbf} shows, for large enough frequencies the usual quadrupole term dominates. Therefore, the actual behavior of the characteristic strain will smoothly interpolate between the scalings described in Eqs.~\eqref{eq:grhc} and \eqref{eq:gbhc}. 
Let us try to have a grasp of when this turnover happens.
Comparing the first and the second term in Eq.~\eqref{eq:gbf}, we find that the dipolar term dominates over the quadrupole when
\begin{equation}
    \frac{5}{96}\frac{\Delta\alpha^2}{x} \frac{q^{2/5}}{\left( 1 + q\right)^{4/5}} \gtrsim 1,
    \label{eq:alphaGB}
\end{equation}
which leads to 
\begin{equation}
    \Delta\alpha^2 \gtrsim 0.01 \frac{(1 + q)^{4/5}}{q^{2/5}}\left(\frac{\mathcal{M}}{10^9 \text{M}_\odot}\right)^{2/3}\left(\frac{f}{10^{-9}\text{Hz}}\right)^{2/3},
\end{equation}
taking for simplicity $G_\text{eff} \sim G$.
Alternatively, the turnover frequency below which the dipole dominates over the quadrupole is 
\begin{equation}
\begin{split}
    f_\text{dom} = &\left(\frac{5}{96}\Delta\alpha^2\right)^{3/2}\frac{\eta^{3/5}}{\pi G_\text{eff}\mathcal{M}}  \\
    \sim &10^{-9}\,\text{Hz}\,\left(\frac{\left|\Delta\alpha\right|}{0.11}\right)^{3}\left(\frac{10^9\text{M}_\odot}{\mathcal{M}}\right)\frac{q^{3/5}}{(1 + q)^{6/5}},
\end{split}
    \label{eq:fdom}
\end{equation}
again taking $G_\text{eff} \sim G$.
The reference values chosen in the previous formulae are particularly informative for a PTA background. Indeed, the recently measured SGWB appears to be dominated by particularly massive ($\mathcal{M}\gtrsim  10^{9}\, \text{M}_\odot$) binaries. Therefore. Eq.~\eqref{eq:alphaGB} gives us a hint about the range of $\alpha$ that could likely produce an effect in the PTA band.\\

\begin{figure}
    \centering
    \includegraphics[width = 0.45\textwidth]{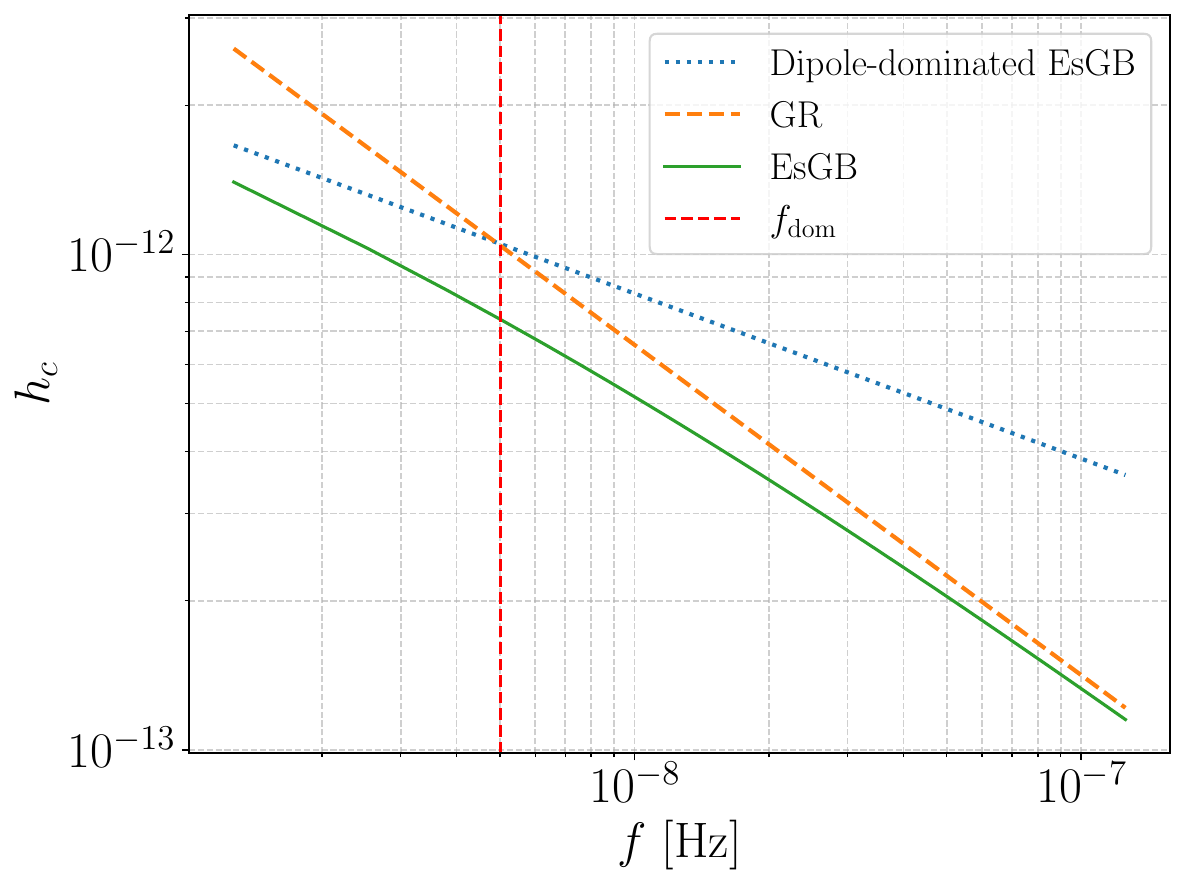}
    \caption{Characteristic strain generated using the merger rate in Eq.~\eqref{eq:mock}, assuming that the background is composed by identical SMBHBs with $\log_{10}\mathcal{\bar M} = 9.6$ and $\bar q = 0.1$, all placed at $\bar z  =1$. The dashed orange line indicates the prediction of GR for circular orbits, see Eq.~\eqref{eq:grhc}; the dotted blue line shows the behavior of the characteristic strain for a dipole-driven inspiral in EsGB theory; the solid green line shows the complete inspiral in the EsGB theory, where we include both the dipolar and the quadrupolar term (see Eq.~\eqref{eq:gbf}). The vertical line shows the expected crossover frequency, determined using Eq.~\eqref{eq:fdom} and appropriately redshifted. We use $\lambda_9 = 5$ to produce this plot, which yields $\Delta\alpha = 0.61$. For the parameters chosen, this curve is practically insensitive to the lower mass cutoff discussed in Sec.~\ref{sec:scacha}.}
    \label{fig:eq}
\end{figure}

In Fig.~\ref{fig:eq}, we show the effects of EsGB gravity on a mock SGWB produced by a population of SMBHBs with the same chirp mass $\log_{10}\mathcal{\bar M} = 9.6\, M_\odot$, the same mass ratio $\log_{10}\bar q = -1$, all placed at the same redshift $\bar z = 1$. In other words, we conveniently simplify Eq.~\eqref{eq:middl} and fix the  merger density to:
\begin{equation}
    \frac{{\rm d}^2n}{{\rm d}z{\rm d}\chi} = 1\,\text{Mpc}^{-3}\delta_D(z - \bar z)\delta_D\left(\log_{10}\frac{\mathcal{M}}{\mathcal{\bar M}}\right)\delta_D\left(\log_{10}\frac{q}{\bar q}\right),
    \label{eq:mock}
\end{equation}
where $\chi = \{\log_{10}\mathcal{M}, \log_{10}q\}$. As expected, the evolution in the early inspiral (dotted blue line) is driven by the dipolar scalar emission, which allows for a faster orbital shrinking than GR. At the turnover frequency, defined by Eq.~\eqref{eq:fdom} and plotted as a red dashed vertical line, the quadrupolar GW emission becomes equal to the dipolar scalar emission, and eventually takes over at larger frequencies.
The resulting characteristic strain (green line), produced using Eq.~\eqref{eq:gbf}, smoothly interpolates between these two different regimes. As a consequence, the spectral index varies continuously from $1/3$ at low frequencies to $2/3$, at the high end of the PTA band. \\

\begin{figure}
    \includegraphics[width = 0.5\textwidth]{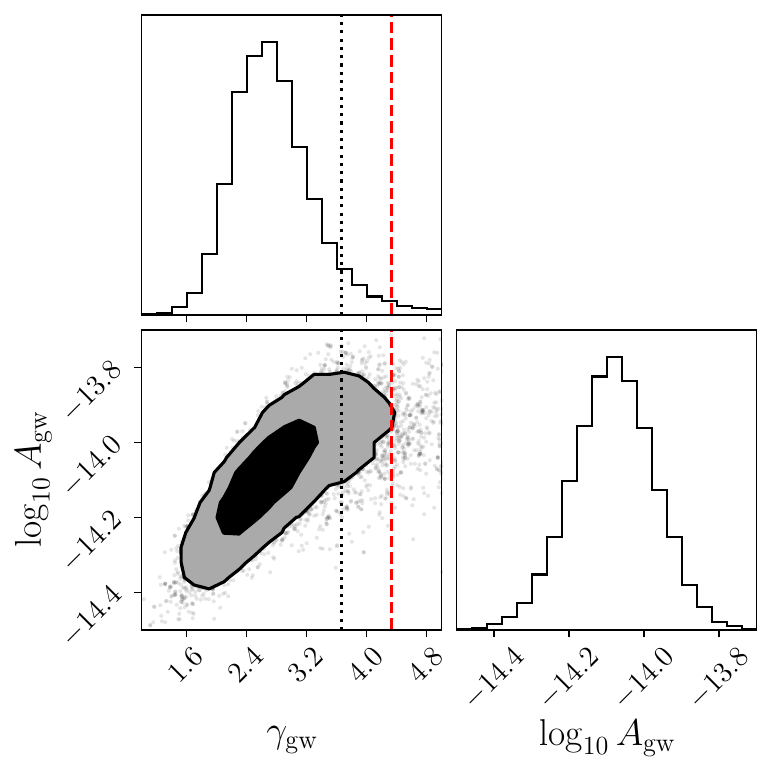}
    \caption{Posteriors on the amplitude and slope of the power spectrum, Eq.~\eqref{eq:pstim}, using the EPTA DR2New dataset, taken from Ref.~\cite{EPTAdata}. The red dashed line shows the spectral index of the timing residuals power spectrum $\gamma_\text{gw}$ (related to the one of the characteristic strain $n$ by $\gamma_\mathrm{gw} = 3 + 2n$, as can be seen from Eq.~\eqref{eq:pstim}) for circularly inspiralling SMBHBs in GR, while the black dotted line shows the prediction in EsGB theory in the dipolar-emission regime, as described in Eq.~\eqref{eq:gbhc}. }
    \label{fig:epta_pl}
\end{figure}

We notice that such a behavior is compatible with recent observations. Indeed, for the parameters used in Fig.~\ref{fig:eq}, EsGB gravity predicts a flatter spectral index than GR at low frequencies. This matches what has been recently observed by PTA collaborations: as an example, in Fig.~\ref{fig:epta_pl}, produced using the EPTA DR2New dataset~\cite{EPTAdata}, we see that the scaling predicted by EsGB in the dipolar-emission regime (black dotted line), as computed by Eq.~\eqref{eq:gbhc}, is visually more in agreement with the data compared to the usual scaling predicted by circularly inspiralling SMBHBs in GR (red dashed line). To produce this plot, we expressed the power spectrum of timing residuals in Eq.~\eqref{eq:pstim} as a power-law, namely
\begin{equation}
    P(f) \equiv \frac{A_\text{gw}^2}{12\pi^2 f_\text{10yr}^3} \left(\frac{f}{f_\text{10yr}}\right)^{-\gamma_\text{gw}},
\end{equation}
normalizing the power spectrum to a pivotal frequency of $f_\text{10yr} = (10\,\text{yr})^{-1}$. Notice that, for a characteristic strain with spectral index $n$, $\gamma_\text{gw} = 3 + 2n$ with our sign conventions. 
See also Fig.~1 and Fig.~10 in Ref.~\cite{Agazie_2023_SGWB} or Fig.~1 and Fig.~5 in Ref.~\cite{EPTA_IV} for further comparisons.\\

However, we must stress that there is a number of different astrophysical effects that could explain a flattening of the spectral index, such as high eccentricities or interaction with the environment~\cite{Enoki_2007, Huerta_2015, Chen:2016kax, Chen_2017, Sesana_2015, Taylor_2017, Burke-Spolaor:2018bvk}. 
While a complete and quantitative comparison among these effects lies beyond the purpose of this work, and could well be the subject of a dedicated analysis, we point out that, in principle, they can still be distinguishable. 
To illustrate this, we follow the analysis of Ref.~\cite{Chen_2017}, which examines orbital shrinking driven by stellar interactions or binary eccentricity. For a circular binary with chirp mass $\mathcal{M}$ and mass ratio $q$, whose inspiral is driven by interactions with stars, the residence time in a log-frequency bin reads
\begin{equation}
    \frac{{\rm d}t_s}{ {\rm d}\log f_s} = \frac{2}{3}\frac{\sigma}{H\rho_i}\frac{q^{1/5}}{(1 + q)^{2/5}}\frac{x}{G^2\mathcal{M}},
    \label{eq:dtstars}
\end{equation}
where $H \sim 15\div 20$ is a dimensionless constant, $\sigma \sim 200 ~\text{km}/\text{s}$ is the typical stellar-bulge velocity dispersion and $\rho_i \sim 100 ~\text{M}_\odot \text{pc}^{-3}$ is the mean stellar density at the binary influence radius, i.e. the radius encompassing a stellar mass twice the total binary mass~\cite{Chen_2017}.~\footnote{Notice the difference in the prefactor compared to Eq.~19 in Ref.~\cite{Chen_2017}. This arises because, while Ref.~\cite{Chen_2017} considers the orbital frequency evolution, here we focus on the GW emission frequency, which for circular binaries is twice the orbital frequency.} 
Combining it with Eq.~\eqref{eq:power} (now setting $\alpha_A, \alpha_B = 0$) and substituting the result in Eq.~\eqref{eq:hc} gives
\begin{equation}
     h_c^2(f) = A_\text{star} f^{2},
\end{equation}
where we omit the explicit form of $A_\text{star}$ for simplicity. Therefore, if the SGWB is dominated by binaries whose orbital evolution is driven by interactions with stars, the spectral index is predicted to be $-1$. However, at sufficiently high frequencies, GW emission is expected to be the primary source of orbital shrinking. Consequently, the shape of the characteristic strain will feature a peak at a frequency $f_\text{peak}$, which can be determined by comparing Eq.~\eqref{eq:dtstars} with Eq.~\eqref{eq:res}. Depending on how strong the interaction with the environment is, then, the spectral index can assume any value between $[-1, 2/3]$.

On the other hand, as we have shown in Sec.~\ref{sec:qual} and in Fig.~\ref{fig:eq}, the EsGB theory of gravity can only yield a spectral index within the range $[1/3, 2/3]$, with the lower bound corresponding to the case where all binaries contributing to the SGWB evolve primarily through EsGB dipolar emission (see Eq.~\eqref{eq:gbhc}). Therefore, if observational data pointed at the presence of a peak in the characteristic strain, rather than a modest flattening of the spectral index, the EsGB hypothesis would be strongly disfavored compared to models involving stellar interactions.

Binary eccentricity is also expected to generate a peak in the characteristic strain. However, as eccentric binaries emit across different frequency bins, calculating the characteristic strain at a specific observing frequency $f$ is less straightforward. Nonetheless, for the purposes of this analysis, the main conclusions remain consistent with those drawn for stellar interactions. We refer the reader to Ref.~\cite{Chen_2017} for a comprehensive treatment of binary eccentricity.
\\

Thus, from a theoretical standpoint, it is possible to distinguish between astrophysical effects - such as interactions with stars and binary eccentricity - and modifications of gravity within the EsGB framework. However, the problem becomes more intricate when computing the characteristic strain produced by a realistic population of SMBHBs: in the EsGB theory, different binaries generally have different turnover frequencies; similarly, for interactions with stars or eccentricity, the location of the peak frequency depends on the underlying properties of the astrophysical population of SMBHBs and their environment.
%Moreover, when computing the characteristic strain produced by a realistic population of SMBHBs, the nice picture developed above becomes more fuzzy, as different binaries generally have different turnover frequencies. 
In any case, the flattening of the spectral index still remains a solid prediction of the EsGB theory, which, for the SGWB under examination, generally yields a value in the range $n \in [ 1/3, 2/3]$. Overall, no conclusive statement can be made at the moment. Future data are expected to improve our understanding of environmental and astrophysical effects and may lead to a final answer.\\

On another note, Fig.~\ref{fig:eq} clearly illustrates that EsGB gravity predicts a smaller amplitude of the SGWB compared to GR at low frequencies. This is easily explained by noticing that the dipolar emission in Eq.~\eqref{eq:gbf} allows systems at low frequency to merge faster. As a result, in order to maintain the merger rate as specified in Eq.~\eqref{eq:mock}, the binaries must be spread across a broader frequency range. As a confirmation, the trend indeed reverses at frequencies higher than $f_\text{dom}$: the dipolar emission becomes less effective, and would lead to slower SMBHB mergers. Consequently, the amplitude of the SGWB would increase, but the quadrupolar emission takes over, and the characteristic strain follows the standard GR prediction.

\section{Results}\label{sec:results}
In this section, we examine the impact of EsGB gravity on PTA data in a more quantitative way. We begin with an idealized setup, generating synthetic timing residuals induced by a nonzero value of the coupling constant of EsGB gravity $\lambda$ and attempting to recover the injected values using a Markov Chain Monte Carlo (MCMC) approach. We then apply the same MCMC analysis to real PTA datasets to assess whether they provide any constraints on EsGB gravity. For convenience, in the following analysis we will express the results in terms of the parameter $\lambda_9$ instead of $\lambda$, as defined below Eq.~\eqref{eq:tildealpha}.

\subsection{Mock injection}\label{sec:idealized}
In order to reduce the possible effects at play, we consider a simplified model for the merger density:
\begin{equation}
\begin{split}
    \frac{d^2n}{{\rm d}z{\rm d}\chi} = &A\left[\left(\frac{\mathcal{M}}{10^7 \text{M}_\odot}\right)^3 \exp\left(-\frac{\mathcal{M}}{10^9\text{M}_\odot}\right)\right] \times\\
    &\times\,p(\log_{10}q)\,\delta_D(z-1 ),
\end{split}
    \label{eq:simp}
\end{equation}
obtained from Eq.~\eqref{eq:middl} fixing $\gamma = -3$, $\mathcal{M}_\star = 10^9\, \text{M}_\odot$ and considering only the mass-dependent term in the distribution, while placing all the SMBHBs at redshift $z = 1$. We assume the usual uniform distribution in $\log_{10}q$ in the range $\log _{10} q \in [-1.5, 0]$ to introduce a a spread in mass ratios among our SMBHBs, thereby varying the relative importance of EsGB gravity compared to GR, as can be seen from Eq.~\eqref{eq:alphaGB}. \footnote{Indeed, note that two equal-mass SMBHBs carry identical scalar charges, resulting in no dipolar emission.}\\

We consider a future PTA experiment, with a timespan of $T_\text{obs} = 30\,\text{yr}$. This sets the minimum frequency and the width of the frequency bins to $f_l = \Delta f \sim 1/T_\text{obs}$. Moreover, we assume we are able to measure the SGWB in the first 20 frequency bins. While this is an idealized setup, it serves to provide a clear illustration of our analysis in a simplified framework. Then, we compute the characteristic strain produced by the astrophysical population in Eq.~\eqref{eq:simp}, fixing the overall normalization to $\log_{10}(A/\text{Mpc}^{-3})= -4$ and injecting a non-zero value of $\log_{10}\lambda_9$ in Eq.~\eqref{eq:hc}, namely $\log_{10}\lambda_9 = 1.9$. In constructing $h_c(f)$, we integrate the chirp mass in the mass range $\log_{10}\mathcal{M}/\text{M}_\odot \in [7, 10.5]$. Chirp masses smaller than the low mass end are expected to contribute negligibly to the signal~\cite{Sesana_2008}, while the high mass end is chosen so as not to produce too massive BHs (for $\mathcal{M} \sim 10^{10.5} \, \text{M}_\odot$ and $\log_{10}q \sim -1.5$, the mass of the primary is $m_1 \sim 2.5\times 10^{11}\, \text{M}_\odot$).  

As the PTA observable is timing residuals, it is customary to express the characteristic strain in terms of the timing residuals induced in the TOAs in a given frequency bin. Looking at Eq.~\eqref{eq:dtdt}, we can write the root-mean-square (rms) of timing residuals in a given frequency bin $i$ as~\cite{EPTA_IV}:
\begin{equation}
\begin{split}
    \delta t_{\text{rms},i} &\sim \left(\int_{\Delta f} P(f) df\right)^{1/2} \sim (P(f_i) \Delta f )^{1/2} \\
    &\sim \left(\frac{P(f_i)}{T_\text{obs}}\right)^{1/2}.
    \label{eq:dti}
\end{split}
\end{equation}
Therefore, after computing the characteristic strain, we construct mock timing residuals as in Eq.~\eqref{eq:dti}. Then, we use Bayesian inference to look for the presence of a signal in the generated mock data. In particular, we leave $\log_{10}(A/\text{Mpc}^{-3})$ and $\log_{10}\lambda_9$ as free parameters and we write the likelihood as 
\begin{equation}
\begin{split}
    \log \mathcal{L}(\log_{10}\delta t|\theta) \propto \sum_i&\left[- \frac{1}{2}\left(\frac{\mu_i-\log_{10}\delta t_i(\theta)}{\sigma_i}\right)^2 \right.\\
    &- \log \left(\sigma_i\sqrt{2\pi}\right)\Bigg] \; ,
    \label{eq:lik}
\end{split}
\end{equation} 
where $\theta \equiv (\log_{10} (A/\text{Mpc}^{-3}), \log_{10}\lambda_9)^T$ and we assumed gaussian measurements, namely
\begin{equation}
    \mu_i = \log_{10}\delta t_\text{rms,i}\,,\quad \sigma_i = \epsilon \mu_i \,,\quad \epsilon = \frac{1}{100}, \frac{1}{50}, \frac{1}{20}\,.
\end{equation}
In the last line, we denoted the uncertainty on the measured value of $\mu_i$ by $\sigma_i = \epsilon \mu_i$. The case $\epsilon = 1/20$ may realistically be achieved in the future for some frequency bins; the other two scenarios are highly idealized (see e.g. Refs.~\cite{EPTA_IV, Agazie_2023_SGWB}). Nonetheless, as mentioned above, they serve to illustrate our analysis in a simplified setting. As priors, we use uninformative ranges, namely $\log_{10}(A/\text{Mpc}^{-3}) \in [-7,0]$ and $\log_{10} \lambda_9 = [-5, 4]$, so as not to bias the recovery.

\begin{figure}
    \centering
    \includegraphics[width = 0.45\textwidth]{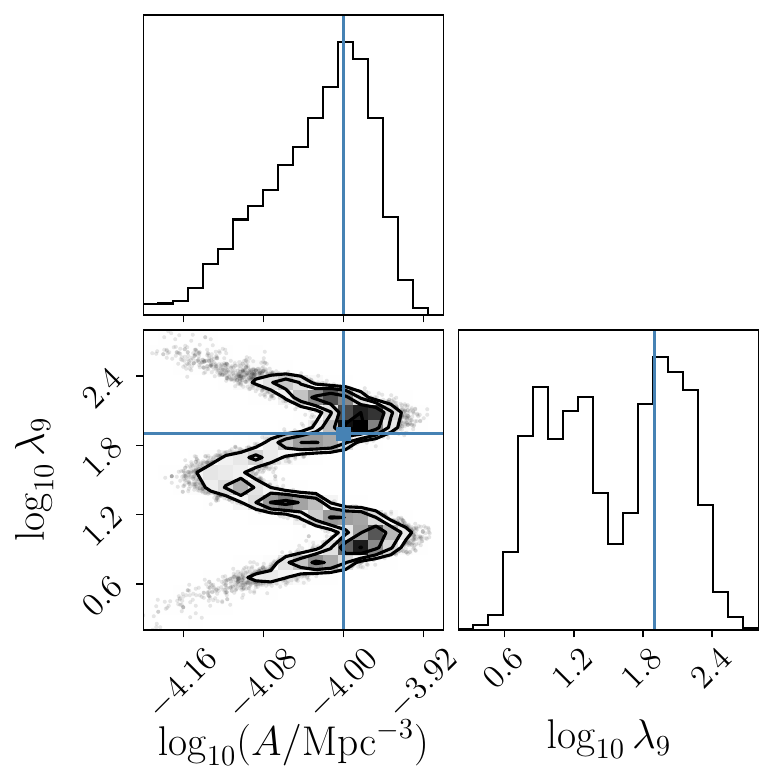}
    \caption{Posteriors on $\log_{10} (A/\text{Mpc}^{-3})$ and $\log_{10} \lambda_9$ for the idealized analysis described in Sec.~\ref{sec:idealized}, setting $\epsilon = 1/100$. The blue solid lines show the injected values.}
    \label{fig:chains_100}
\end{figure}

In Fig.~\ref{fig:chains_100}, we plot the result of the Bayesian inference for $\epsilon = 1/100$, while we display the results for $\epsilon = 1/50, 1/20$ in Appendix~\ref{app:other_plots}. We find that the amplitude of the injected merger density is quite well constrained.  In contrast, while the parameter $\log_{10}\lambda_9$ is clearly detected in the data, its posterior distribution is bimodal and spans several orders of magnitude—even within this idealized scenario.

However, the shape of the 2D corner plot elucidates the nature of the degeneracy and, to some extent, clarifies why we are fundamentally limited in precisely extracting the true value of $\lambda_9$.
Indeed, let us remark that, in the theory we are considering, a given value of $\lambda_9$ predicts a window of scalarized masses, as shown in Fig.~\ref{fig:b5}. As $\lambda_9$ increases (decreases), the window shifts to higher (lower) BH masses, maintaining its amplitude and width (in log space) constant.
Therefore, considering a system of two SMBHBs with different masses, there is a degeneracy: in other words, as the frequency evolution in the dipolar-dominated regime is only concerned with the absolute difference of scalar charges (at least at lowest PN order, see Eq.~\eqref{eq:gbf}), we cannot distinguish the situation in which the window is centered on one of the two masses from the opposite one. 
As a consequence, the constraints on $\log_{10} \lambda_9$ will manifest this degeneracy, as clearly shown in Fig.~\ref{fig:chains_100}.\\

\begin{figure}
    \centering
    \begin{subfigure}[b]{0.45\textwidth}
        \centering
        \includegraphics[width=\textwidth]{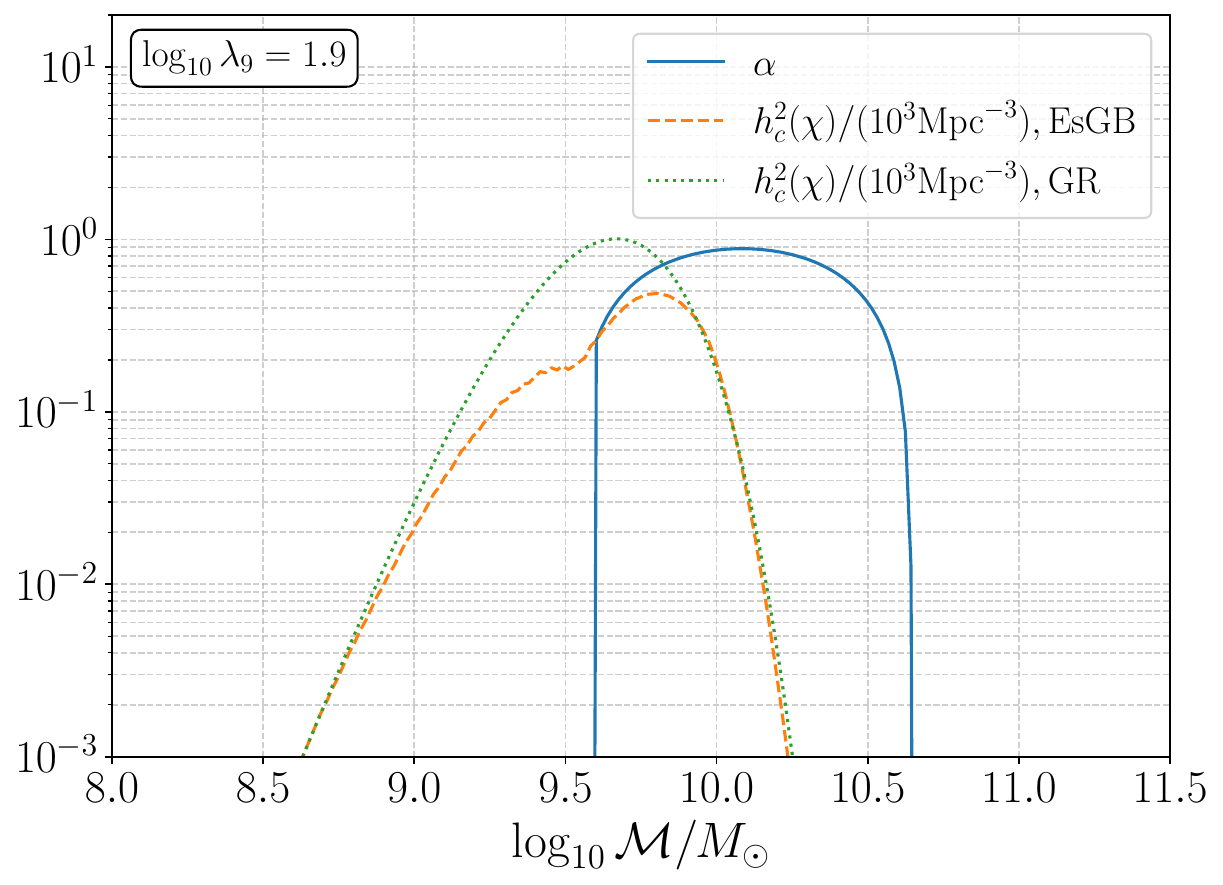}
    \end{subfigure}
    \hfill
    \begin{subfigure}[b]{0.45\textwidth}
        \centering
        \includegraphics[width=\textwidth]{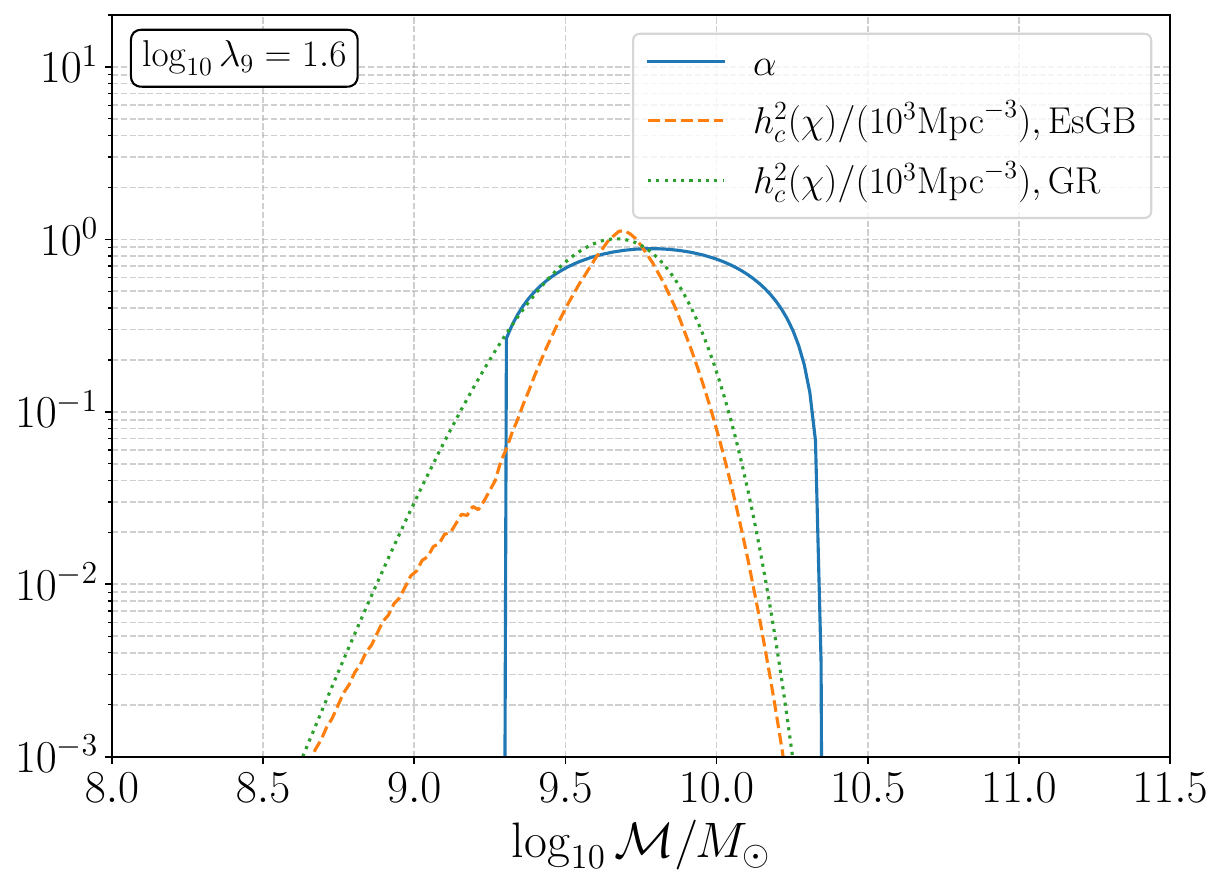}
    \end{subfigure}
    \begin{subfigure}[b]{0.45\textwidth}
        \centering
        \includegraphics[width=\textwidth]{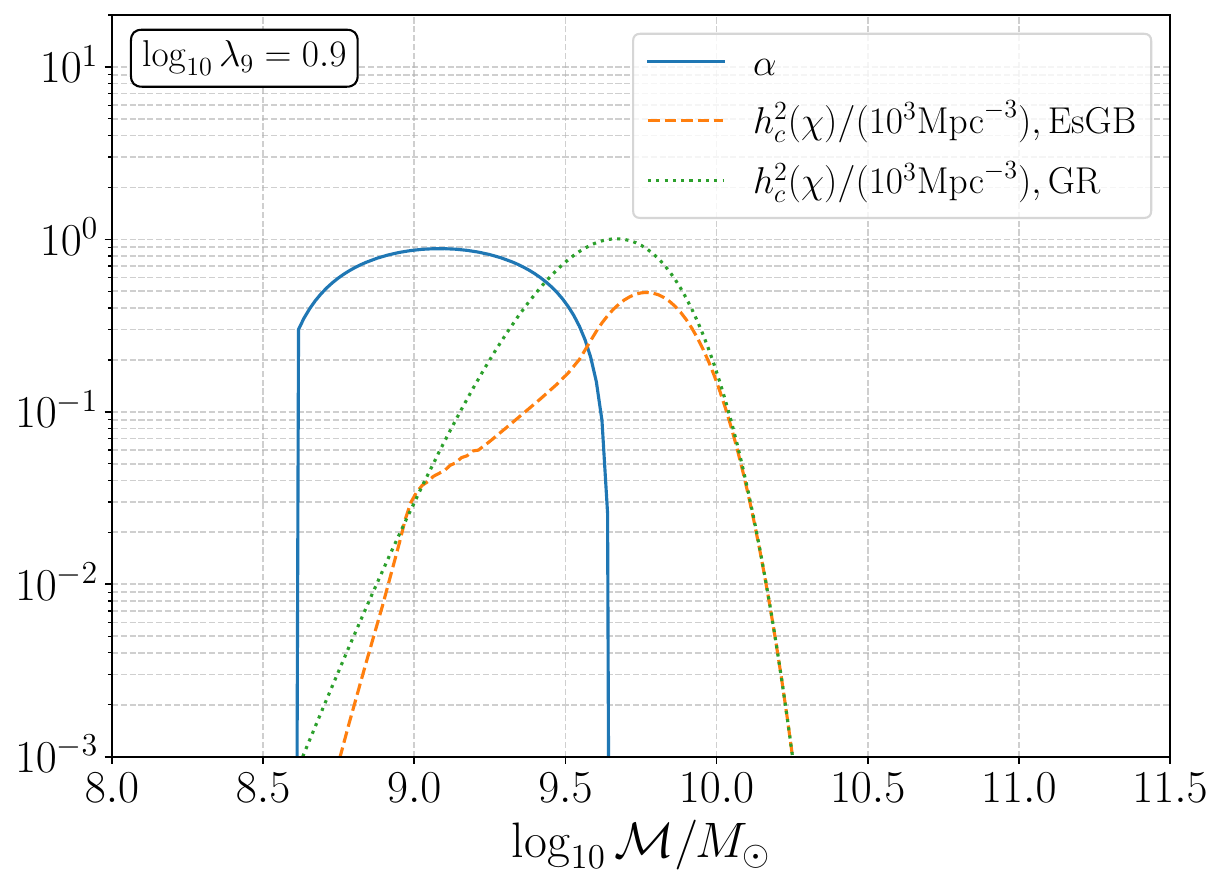}
    \end{subfigure}
    \caption{Scalarized window of SMBH masses and \textit{source loudness}, as defined in Eq.~\eqref{eq:loud} for $f = 10^{-9} \text{Hz}$. The curves are produced assuming a uniform distribution in $\log_{10}q$ ranging from $\log_{10}q    \in [-1.5, 0]$, and assuming the merger density in Eq.~\eqref{eq:simp}. The top panel displays the scalarization window for $\log_{10} \lambda_9 = 1.9$ and the resulting \textit{source loudness} in EsGB theory. The middle and low panels show the same curves, but for $\log_{10} \lambda_9 = 1.6$ and $\log_{10} \lambda_9 = 0.9$  . In the three panels, we also overplot the \textit{source loudness} in GR, which does not depend neither on the scalarization window nor on the mass ratios of the systems.}
    \label{fig:loudness}
\end{figure}

As a matter of fact, such a degeneracy can be partially broken by the astrophysical population. To gain more insight, let us start by rewriting
Eq.~\eqref{eq:hc} as 
\begin{equation}
    h_c^2(f) = \frac{4G}{\pi f^2}\int {\rm d}\chi\,  h_c^2(\chi, f),
    \label{eq:loud}
\end{equation} 
where the \textit{source loudness} $h_c^2(\chi, f)$ can be found by simple substitution of Eqs.~\eqref{eq:power} and \eqref{eq:gbf} in Eq.~\eqref{eq:hc}, assuming that $z = 1$ for all the SMBHBs.
In Fig.~\ref{fig:loudness}, we plot $h_c^2(\chi, f)$ for GR and EsGB gravity theories, using Eq.~\eqref{eq:res} and Eq.~\eqref{eq:gbf}, respectively, fixing a benchmark frequency of $f = 10^{-9}\, \text{Hz}$ and assuming the merger density in Eq.~\eqref{eq:simp}.  Moreover, we overplot the scalarized window of SMBH masses resulting for three representative values, namely $\log_{10}\lambda_9 = 1.9, 1.6, 0.9$. These values are roughly located in both the two peaks and the valley in Fig.~\ref{fig:chains_100}. To produce the EsGB curves in Fig.~\ref{fig:loudness}, we assume a log-uniform prior in the mass ratio characterizing the SMBHBs, in the interval $\log_{10} q\in [-1.5,0]$. We now provide a more detailed explanation of the shape of the posterior shown in Fig.~\ref{fig:chains_100}.
\\

As defined in Eq.~\eqref{eq:loud}, the source loudness is a measure of the contribution of SMBHB sources with parameter $\chi$ to the total SGWB. The peak of the source loudness defines the properties of the SMBHB systems which contribute the most to the SGWB.
Firstly, we notice that the presence of EsGB gravity generally decreases the source loudness of SMBHBs, resulting in a lower amplitude of the SGWB, as already expected from Fig.~\ref{fig:eq}.\footnote{An exception is found in the middle panel of Fig.~\ref{fig:loudness}, where the peak source loudness in EsGB gravity exceeds the corresponding GR value. This occurs because, when the scalar charges of the two SMBHs in a binary are nearly equal, the orbital decay is primarily driven by the quadrupole component, as indicated by Eq.~\eqref{eq:gbf}. In this scenario, the characteristic strain in EsGB gravity is enhanced by an overall factor of $(1 + \alpha_A\alpha_B)^{2/3}$ compared to the GR case, as can be shown by substituting Eqs.~\eqref{eq:power} and \eqref{eq:gbf} into Eq.~\eqref{eq:dEdf}, and subsequently into Eq.~\eqref{eq:hc}. The resulting SGWB, which is the integral of the source loudness over the parameters $\chi$, remains however lower than the GR counterpart, even in this scenario.}

Secondly, comparing the upper and the lower panels in Fig.~\ref{fig:loudness}, we find that the interval of chirp masses contributing the most to the background is roughly the same in these two cases, and also the relative source loudness is comparable. In other words, these two panels show the scenario in which the scalarization window is centered around only one of the two bodies forming a typical SMBHB system in our model. As previously mentioned, this is the source of the degeneracy shown in the 2D posterior in Fig.~\ref{fig:chains_100}. Such a degeneracy is not exact, because of the specifics of the underlying astrophysical model described by the comoving merger density: indeed, we see that the source loudness is not exactly the same in the two cases, which is reflected in the different shape of the two peaks in the 2D posterior. However, such a breaking is not sufficiently strong to disentangle the two cases in the data. 

Finally, we observe the presence of a valley between the two peaks of the bimodal distribution in Fig.~\ref{fig:chains_100}. As shown in the middle panel of Fig.~\ref{fig:loudness}, this feature arises because, in that region of parameter space, most SMBHB systems consist of two scalarized black holes with similar values of the adimensional charge $\alpha$. In such cases, the charge difference is generally small, leading to weaker dipolar emission compared to scenarios where only one of the two black holes is scalarized. Consequently, the peak of the source loudness in EsGB gravity is not significantly modified compared to GR, implying that the overall shape of the SGWB remains largely unchanged.
This results in a suppression of the posterior in the corresponding region of $\log_{10} \lambda_9$.

Nonetheless, a non-negligible number of posterior samples still populate this region, and they tend to be correlated with a lower amplitude for the merger rate.  This is ultimately caused by the fact that, as we have shown in Figs.~\ref{fig:eq} and \ref{fig:loudness}, the presence of EsGB gravity suppresses the overall amplitude of the SGWB. 
Therefore, the mock data points we constructed can also be fitted - although with a smaller likelihood - by a smaller $\log_{10} (A/\text{Mpc}^{-3}) $ and values of $\log_{10} \lambda_9$ which do not produce a heavy suppression of the SGWB.
As a consequence, our analysis cannot fully exclude the values of $\log_{10} \lambda_9$ in the trough of Fig.~\ref{fig:chains_100}, although strongly disfavoring them. \\

The last point becomes increasingly relevant as the precision of our mock measurements decreases - that is, if $\epsilon$ increases. In Appendix~\ref{app:other_plots} (Fig.~\ref{fig:app_chains_5020}), we show the results for $\epsilon = 1/50, 1/20$. In the former case, the GR fit is more in agreement with the data compared to the $\epsilon = 1/100$ case: the posterior in  $\log_{10}(A/\text{Mpc}^{-3})$ becomes bimodal - signaling that the data could be fitted by pure GR, although with a smaller likelihood - while the through in $\log_{10} \lambda_9$ disappears. In the latter case, the uncertainty in the mock measurements is so large that even the trough in the amplitude disappears, resulting in significantly broader constraints. However, the characteristic structure in the 2D corner plot remains discernible and continues to serve as a distinctive imprint of EsGB gravity on the data.

We checked that choosing different parameters in Eq.~\eqref{eq:simp} yields qualitatively similar results.

\subsection{Application to real data}\label{sec:realPTA}
Here, we apply the analysis outlined in the section above to real data. Specifically, we focus on the EPTA DR2New dataset, which consists of the most recent 10.3 years of observations from the full EPTA DR2 dataset (spanning 24.7 years), obtained using new-generation backend systems.

We construct the characteristic strain as in Eq.~\eqref{eq:hc}, using the agnostic comoving merger density in Eq.~\eqref{eq:middl}. Then, we compute the induced timing residuals $\delta t_{\text{rms}, i}$ as in Eq.~\eqref{eq:dti}, where $P(f)$ is defined by Eq.~\eqref{eq:pstim}, comparing them to the violin plots presented in Ref.~\cite{EPTA_IV}. Finally, we set up a Bayesian MCMC search for the parameters defining the characteristic strain, namely $\{\log_{10} \frac{\dot{n}_0}{\text{Mpc}^{-3}\text{Gyr}^{-1}}, \log_{10} \lambda_9, \beta, z_0, \gamma, \log_{10}\mathcal{M}_\star/\text{M}_\odot\}$. We define the log likelihood as 
\begin{equation}
    \log \mathcal{L}(\log_{10}\delta t|\theta) \propto \sum_i^{N_\text{bins}}\text{Prob}_i\left(\log_{10}\delta t_i(\theta)\right),
    \label{eq:lik_EPTAnew}
\end{equation}
where $\text{Prob}_i$ is the normalized probability distribution constructed from the $i$-th bin in Fig.~1 of Ref.~\cite{EPTA_IV}. We consider $N_\text{bins} = 9$, as commonly done in the literature. However, we point out that this might introduce some bias in the analysis, as spikes from single sources dominating over the background are expected to show up at frequencies larger than $f\gtrsim 10^{-8}\, \text{Hz}$~\cite{Sesana_2008}. In this case, the parametric form in Eq.~\eqref{eq:hc} with the merger rate distribution in Eq.~\eqref{eq:middl} is not expected to hold anymore, and the distribution of the resulting characteristic strain around the value in Eq.~\eqref{eq:hc} may exhibit significant scatter, due to the particular realization of the Universe~\cite{Bonetti_2024}.
To account for this, in Appendix~\ref{app:other_plots} (Fig.~\ref{fig:EPTADR2New_4}) we present  an analysis performed on the first four frequency bins, which are expected to be well within the region of validity of the description in Eqs.~\eqref{eq:hc} and \eqref{eq:middl}.\\

\begin{figure}[H]
    \centering
    \includegraphics[width = \textwidth]{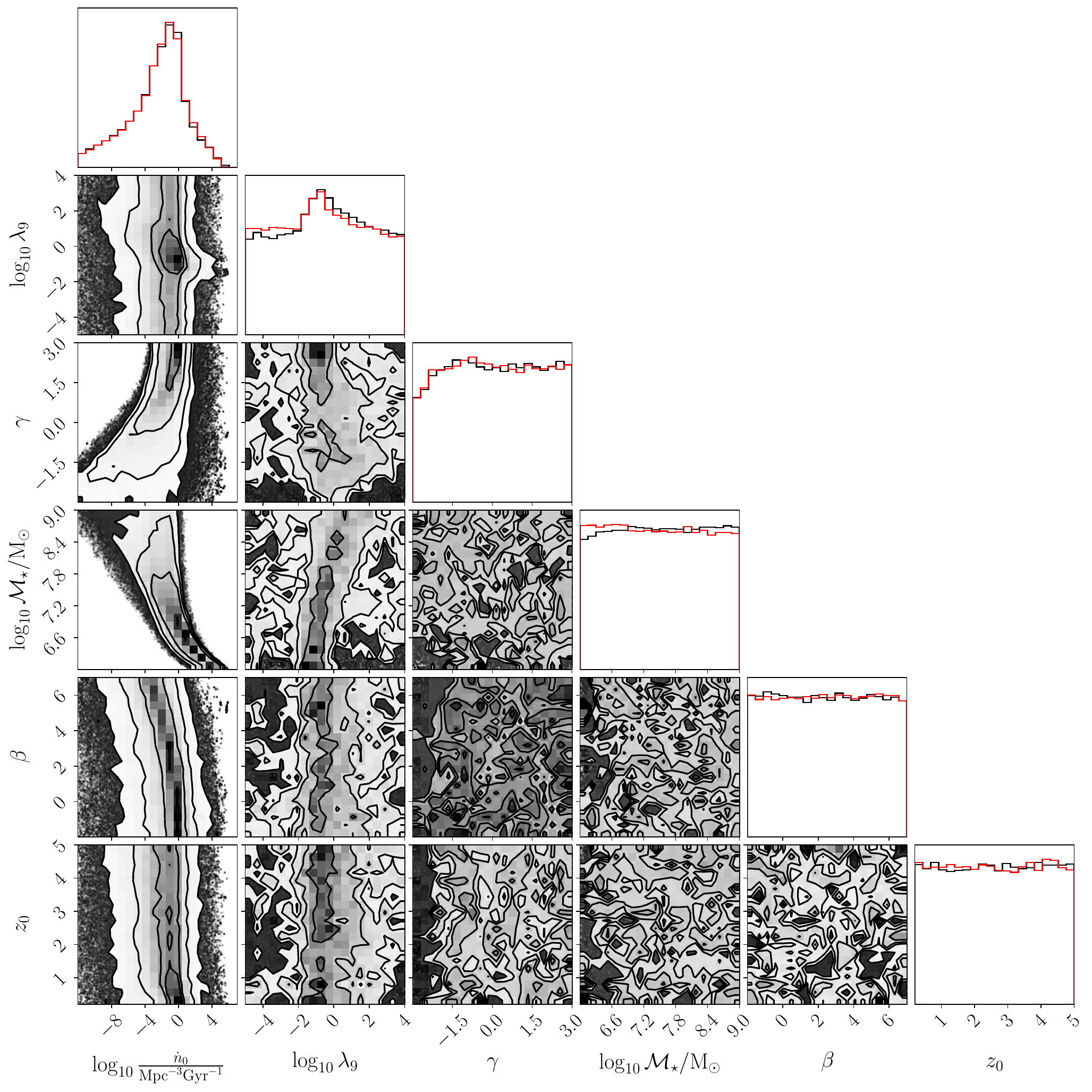}
    \caption{Posteriors on $\{\log_{10} \frac{\dot{n}_0}{\text{Mpc}^{-3}\text{Gyr}^{-1}}, \log_{10} \lambda_9, \gamma,  \log_{10}\mathcal{M}_\star/\text{M}_\odot, \beta, z_0\}$ for the analysis of the EPTA DR2New data described in Sec.~\ref{sec:realPTA}, using the first nine frequency bins. The black contours label the posteriors for the case in which $\log_{10}q \in [-1.5,0]$. The red marginals overplot the posteriors when a uniform distribution in the same interval of mass ratios, namely $ q \in [0.03, 1]$, is assumed. }
    \label{fig:EPTADR2New_9}
\end{figure}

As done in Sec.~\ref{sec:idealized}, we integrate the chirp mass in the range $\log_{10}\mathcal{M}/\text{M}_\odot \in [7, 10.5]$ and the redshift $z$ in the range $z  \in [0, 5]$. The priors on the parameters of the search are taken from Sec.~\ref{sec:com}, and are chosen to be consistent with Refs.~\cite{ Middleton_2015, EPTA_IV}. Furthermore, we sample $\log_{10}\lambda_9$ in the range $\log_{10}\lambda_9 \in [-5,4]$, which largely encompasses all the possible values of $\lambda_9$ that may scalarize SMBHBs in the PTA mass range.
We present the results of the analysis in Fig.~\ref{fig:EPTADR2New_9}, while we show the results of a similar analysis of the first 14 frequency bins the NANOGrav 15yr dataset~\cite{Agazie_2023_tim, Agazie_2023_SGWB, nanodata} to Fig.~\ref{fig:NANOGrav} in Appendix~\ref{app:other_plots}.\\

The amplitude of the comoving merger density ($\log_{10} \frac{\dot{n}_0}{\text{Mpc}^{-3}\text{Gyr}^{-1}}$) is well constrained, whereas the remaining parameters largely reflect the prior. We report a mild excess in the posterior of $\log_{10}\lambda_9$ in the region $-2\lesssim \log_{10}\lambda_9 \lesssim 0$, which implies a slight preference for $1.5\times10^7~\text{km}\lesssim \lambda\lesssim 1.5\times10^9~\text{km}$. However, the Bayes factor comparing EsGB to GR is only $\sim1.35$, suggesting that this preference is not statistically significant (see Appendix~\ref{app:bayes} for details on the Bayes factor calculation).
Therefore, although we report a small indication for $1.5\times10^7~\text{km}\lesssim \lambda\lesssim 1.5\times 10^9~\text{km}$, current PTA data are not sensitive enough yet to conclusively discriminate between the presence of EsGB gravity and GR. This is reflected in the marginalized posterior of $\log_{10}\lambda_9$, which more closely resembles the right panel of Fig.~\ref{fig:app_chains_5020} in the Appendix than Fig.~\ref{fig:chains_100}. 

Qualitatively, values of $\log_{10}\lambda_9$ sufficiently above or below the peak yield a SGWB spectrum that is indistinguishable from that predicted by GR, as the scalarization window shifts to masses outside the range contributing to the PTA signal (see also Fig.~\ref{fig:loudness} for a visual comparison). This ultimately explains why, if the data are not precise enough to disentangle EsGB gravity from GR, the posterior for $\log_{10}\lambda_9$ shows no fall-off behavior at either large or small values (as opposed to what happens, for instance, in the idealized case of Fig.~\ref{fig:chains_100}). 

The enhancement is even less pronounced in the 4-bins analysis in Fig.~\ref{fig:EPTADR2New_4} in Appendix~\ref{app:other_plots} (Bayes factor $\sim 1.25$), and, in general, the posteriors remain too broad to allow for the identification of a distinctive shape in the 2D corner plot of $\{\log_{10}\frac{\dot{n}_0}{\text{Mpc}^{-3}\text{Gyr}^{-1}}, \log_{10}\lambda_9\}$, such as the one in Fig.~\ref{fig:chains_100}.
Improved data quality and a better understanding of the underlying astrophysical SMBHB population could help determine whether this excess is a mere statistical fluctuation or a genuine hint for a fundamental physical process.\\

In Fig.~\ref{fig:EPTADR2New_9}, we examine two reference distributions for the mass ratio $q$: a log-uniform distribution over $\log_{10}q\in[-1.5, 0]$ (solid black lines) and a uniform distribution in the same range, i.e. $q \in [0.03,1]$ (solid red lines). The resulting posteriors do not show any appreciable difference, and we present only the former choice in Figs.~\ref{fig:EPTADR2New_4} and \ref{fig:NANOGrav} in Appendix~\ref{app:other_plots}. This qualitatively illustrates that our findings are fairly robust against variations in the mass ratio distribution, provided they are not strongly peaked around $q=1$, in which case the effect of EsGB gravity would be suppressed, as already pointed out in Sec.~\ref{sec:com}.
\onecolumngrid

\twocolumngrid
\clearpage
\newpage
Should the peak in the $\log_{10}\lambda_9$ posterior be confirmed, it would still be essential to rule out alternative explanations, such as strong eccentricity or significant interaction with the environment, before interpreting it as evidence for EsGB gravity. As discussed earlier, while these effects present characteristic distinguishable signatures, they may be degenerate to some extent when considering a realistic SMBHB population.
%As discussed earlier, these effects are largely degenerate.
Future measurements may help break this degeneracy and clarify the origin of the observed features.
\section{Conclusions}\label{sec:concl}

ST theories are  well-motivated extensions of GR, characterized by the presence of at least one scalar DOF in the gravitational sector. 
Although no-hair theorems ~\cite{Bekenstein:1971hc, Bekenstein:1972ky,Graham:2014mda,Hui:2012qt,Capuano:2023yyh} typically prevent ST theories from modifying black hole solutions at the background level, it is possible to envision classes of theories in which such theorems are circumvented~\cite{Sotiriou:2013qea,Sotiriou:2014pfa,Doneva:2017bvd,Doneva:2021tvn,Eichhorn:2023iab}. 
In this work, we analyzed the EsGB theory of gravity, reviewed in Sec.~\ref{sec:AH}, focusing on the specific model given by Eq.~\eqref{Aaron_model}. Unlike standard EsGB, this model predicts a non-trivial scalar hair around BHs in a well-delimited range of masses, allowing for scalarization to occur only at supermassive scales. \\

For this reason, PTAs stand out as uniquely suited experiments to test this theory. 
PTAs monitor a network of millisecond pulsars over decade-long experiment. The recent detection of a temporally-correlated common stochastic noise, featuring a distinctive correlation among pulsars TOAs, has opened a window not only into astrophysical phenomena, but also potential signature of new physics (see e.g. Refs~\cite{Sanidas_2012, Khmelnitsky_2014, Caprini_2018, Blasi_2021, Chen_2022, Kaplan:2022lmz, EPTA_quelq, Smarra:2023egv, Kim:2023kyy, Cannizzaro:2023mgc, Frosina:2023nxu, EPTA_IV, Afzal_2023, Figueroa:2023zhu, Ellis:2023oxs, Madge:2023dxc, Gouttenoire:2023ftk, Liu_2023, Chen:2024fir, Ellis_2023,  Franciolini:2023wjm, Franciolini:2023pbf, Smarra:2024kvv, porayko2024, wu2024, Bi_2024, Athron_2025} and references therein).

In this work, we assume that the recently observed common excess in pulsars' TOAs arises from the incoherent superposition of GW emitted by circularly inspiralling SMBHBs. We discuss how the specific variant of EsGB gravity from Ref.~\cite{Eichhorn:2023iab} and described in Sec.~\ref{sec:AH} impacts the characteristic strain generated by the underlying astrophysical population. 
These effects are illustrated in Fig.~\ref{fig:eq} %\AK{rather Fig.~\ref{fig:epta_pl} no?}
, which shows that this model leads to a flattening of the spectral index compared to the prediction from circular binaries in pure GR. Notably, this behavior aligns with current PTA observations, as shown in Fig.~\ref{fig:epta_pl} for the EPTA DR2New dataset.
Furthermore, as the EsGB theory of gravity predicts a regime dominated by dipolar emission for binaries involving at least one scalarized SMBH, we also observe a decrease in the overall amplitude of the characteristic strain with respect to GR, keeping the parameters in the merger density fixed. 
It may be interesting to confront this prediction with astrophysically-motivated priors on the merger density parameters, as done in Refs.~\cite{EPTA_IV, Chen_2019}. In particular, EsGB gravity may shift the amplitude toward the high end of the prior range, leading to a slight tension. We leave this aspect for future work.\\

After analyzing the main features of EsGB gravity using mock data in Sec.~\ref{sec:idealized}, we consider real PTA observations in Sec.~\ref{sec:realPTA}. In the main text, we report results for the EPTA DR2New dataset, where we find a  marginal preference for values of $\lambda$  in the range $1.5\times10^7~\text{km}\lesssim \lambda\lesssim 1.5\times10^9~\text{km}$. However, it is important to emphasize that several astrophysical mechanisms, as high eccentricities or interaction with the environment~\cite{Enoki_2007, Huerta_2015, Chen:2016kax, Chen_2017, Sesana_2015, Taylor_2017, Burke-Spolaor:2018bvk}, could also lead to a flattening of the spectral index, making them largely degenerate with the signatures expected from EsGB gravity.
Future data releases, with longer observation times and enhanced precision, will definitely advance our comprehension of the astrophysical processes influencing the SGWB produced by SMBHBs. These advancements may help disentangle the different effects and ultimately lead to a final conclusion.

\section{Acknowledgements}
We warmly thank Alberto Sesana for insightful discussions and important suggestions. Moreover, we wish to thank Enrico Barausse and Aaron Held for helpful comments and discussions on the manuscript, Nataliya Porayko for kindly providing a code to reproduce the free spectrum of timing residuals for the EPTA dataset and Rahul Srinivasan for useful discussions about Bayesian inference. L.C. acknowledges support from the European Union’s H2020 ERC Consolidator Grant ``GRavity from Astrophysical to Microscopic Scales'' (Grant No. GRAMS-815673),  the European Union’s Horizon ERC Synergy Grant ``Making Sense of the Unexpected in the Gravitational-Wave Sky'' (Grant No. GWSky-101167314), the PRIN 2022 grant ``GUVIRP - Gravity tests in the UltraViolet and InfraRed with Pulsar timing'', and the EU Horizon 2020 Research and Innovation Programme under the Marie Sklodowska-Curie Grant Agreement No. 101007855. A.K. acknowledges funding from the FCT project ``Gravitational waves as a new probe of fundamental physics and astrophysics'' Grant Agreement No. 2023.07357.CEECIND/CP2830/CT0003.

\appendix

\section{BINARY DINAMICS IN THE POST-NEWTONIAN APPROXIMATION} \label{app:binary_dynamics}
In this section, we want to provide a schematic picture of the derivation of the equations for the evolution of the orbital frequency that we employed in the main text. Here we summarize the main steps; for a more in-depth discussion, the reader is referred to Refs.~\cite{Juli__2019,Shiralilou:2021mfl}. 

In ST theories, a binary system of bodies carrying a scalar charge can be modeled by supplementing the vacuum action with a contribution $S_m$ describing two point particles, that we label with the subscripts $A$ and $B$. Namely,
\begin{equation}
    S_m = -\int{\rm d}s_A\,m_A(\varphi)+(A\leftrightarrow B)\,,
\end{equation}
where ${\rm d}s_A=\sqrt{-g_{\mu\nu}{\rm d}X_A^\mu{\rm d}X_A^\nu}$, and $X^\mu_A[s_A]$ represents the world-line of the point-particle $A$.

Despite the notation, it is important to stress that $S_m$ is not a proper matter action, as the point particles represent here \textit{skeletonized} BHs, which can be matched to vacuum solutions of the field equations in the large-distance limit. In particular, the scalar field is coupled to the point particles through $\varphi$-dependent masses, encoding the fundamental coupling of the scalar field to the gravitational sector. On the other hand, we consider the coupling between matter and the scalar field to be trivial.
This picture can be connected to the scalar charge and mass defined in the main text through the matching conditions~\cite{Juli__2019}
\begin{equation}
    \begin{cases}
    &M_A = m_A(0)\\
    &\alpha_A=-\frac{{\rm d}\log m_A}{{\rm d}\varphi}\left.\right|_{\varphi=0}\,.
\end{cases}
\end{equation}
Varying the total action, given by $S_{\rm EsGB}+S_m$, with respect to the metric and the scalar field, we find the EOMs 
\begin{equation}
\begin{split}
    G_{\mu\nu}=& 2 \partial_\mu\varphi\partial_\nu\varphi -  g_{\mu \nu} \left( (\partial\varphi)^2 - \frac{\lambda^2}{2} F(\varphi) f(\mathcal G )\right) \\
&+ 4 \lambda^2 \vphantom{}^{**} R_\mu{}^{\alpha \beta}{}_\nu\nabla_\alpha\nabla_\beta\left(F(\varphi)f'(\mathcal G)\right)+\\
&+8\pi G T_{\mu\nu}^A+(A\leftrightarrow B)\,, \\
\Box\varphi = & \frac{\lambda^2}{4}F'(\varphi)f(\mathcal G)\\
&+4\pi G \int \mathrm{d} s_A \frac{{\rm d}m_A}{{\rm d}\varphi}\frac{\delta^{(4)}(x-x_A(s_A) )}{\sqrt{-g}}\\&+(A\leftrightarrow B)\, ,
\end{split}
\label{full_system}
\end{equation}
where $G_{\mu\nu}$ and $\vphantom{}^{**}R_{\rho\alpha\beta\sigma}$ are respectively the Einstein tensor and the double-dual Riemann tensor, and we defined the stress energy tensor associated with a point-particle as
\begin{equation}
    T^{\mu\nu}_A=\int {\rm d}s_A\,m_A(\varphi)\frac{\delta_D^{(4)}(x-x_A(s_A))}{\sqrt{-g}}\frac{{\rm d}X_A^\mu}{{\rm d}s_A}\frac{{\rm d}X_A^\nu}{{\rm d}s_A}\,.
\end{equation}
Let us now consider a binary system of BHs.
In what follows, we will define the Keplerian orbital velocity as 
\begin{equation}
    v =\sqrt{\frac{G_{\text{eff}}M}{r}}\,,
    \label{eq:KeplerLaw}
\end{equation}
where $r$ is the separation between the two bodies, $M = M_A+M_B$ is the total mass, and $G_{\text{eff}}=(1+\alpha_A\alpha_B)G$ is the effective gravitational coupling constant accounting for the effect of scalar charges.

We assume that the orbits of the two BHs during the early inspiral are quasi-circular and Keplerian, namely that their orbital velocity is always given by Eq.~\eqref{eq:KeplerLaw} and $\dot{r}\simeq 0$.
The metric in the Near Zone can be expanded around the flat Minkowski spacetime as done in Ref.~\cite{Juli__2019} 
\begin{equation}
\begin{split}
    &g_{00}=\exp{\left(-2U\right)}+\mathcal{O}(v^6)\,,\\
    &g_{i0}=-4 g_i+\mathcal{O}(v^5)\,,\\
    &g_{ij}=\delta_{ij}\exp{\left(2U\right)}+\mathcal{O}(v^6)\,,
\end{split}
\label{eq:expans}
\end{equation}
with $U\sim\delta\varphi\sim\mathcal{O}(v^2)$ and $g_i\sim \mathcal{O}(v^3)$.
On the other hand, the scalar field and the gravitational masses can be expanded as
\begin{equation}
\begin{split}
    &\varphi=\varphi_0+\delta\varphi+\mathcal{O}(v^6)\,,\\
    &m_A(\varphi) = M_A\left(1-\alpha_A\delta\varphi+\left((\alpha_A)^2-\alpha_A'\right)\delta\varphi^2\right)\\
    &\qquad\qquad+\mathcal{O}(v^6)\,,
\end{split}
\end{equation}
where we defined $\alpha'_A={\rm d}\ln\alpha_A/{\rm d}\varphi\left.\right|_{\varphi=0}$.
The BHs are then fully determined, modulo next to leading order corrections, by their mass, their scalar charges and the first derivatives of the scalar charges computed at $\varphi = 0$.\\
Using Eq.\eqref{eq:expans}, the Gauss-Bonnet invariant then reads
\begin{equation}
\begin{split}
    \mathcal{G}=&8\left(\partial_i\partial_j U\partial^i\partial^j U-\nabla^2U \nabla^2 U)\right)+\mathcal{O}(v^6)\\
    &\sim \mathcal{O}(v^4)\,.
\end{split}
\end{equation}
Given the form of the EOMs \eqref{full_system}, the terms proportional to $\mathcal{G}^n$, with $n\geq 2$ come therefore with a further $v^{4(n-1)}$ suppression, so we will ignore them in the following. This amounts to considering the case $f(\mathcal{G})\simeq -\mathcal{G}$.

Employing the PN expansion of the relevant fields, together with Eq.~\eqref{full_system}, one can obtain the binding energy $E_B$ of the system in terms of the properties of the two bodies. At leading order, we have
\begin{equation}
    E_B=-\frac{\mathcal{M}x}{2}+\mathcal{O}(x^2)\,,
\end{equation}
where we introduced the standard PN expansion parameter
\begin{equation}
    x\equiv \left(\pi G_{\text{eff}}\mathcal{M}f_s\right)^{2/3}
\end{equation}
with $f_s$ being the GW emission frequency in the source frame. The parameter $x$ is related to $v$ via $x = q^{2/5}/(1+q)^{4/5} v^2$.
Within the same framework, it is also possible to derive the power emitted through the different channels using a matching of the metric with a far-zone solution~\cite{Shiralilou:2021mfl}. We report here only the final expressions.
The power radiated through the tensor channel reads
\begin{equation}
    \dot{E}_{\text{T}}= \frac{32}{5 G_\mathrm{eff} (1+\alpha_A\alpha_B)} x^5+\mathcal{O}\left(x^6\right)\,.
\end{equation}
Conversely, the power emitted through the scalar channel is
\begin{equation} \label{eq:EdotS}
\begin{split}
\dot E_S=&\frac{x^4}{3 G_{{\rm eff}}(1+\alpha_A\alpha_B)}\left[\eta^{2/5}\Delta\alpha^2\left(1+\mathcal{O}(x)\right)\right.\\
&\left.+x\left(\frac{4}{5}(\alpha_A+\alpha_B)^2+W+Q_\mathcal{G}\right)\right]\\
&+\mathcal{O}\left(x^6\right)\,.
\end{split}
\end{equation}
The first term, proportional to the squared difference of the scalar charges $\Delta \alpha^2$, corresponds to dipolar emission.
Notice that $\Delta\alpha^2$ also multiplies terms of order $\sim\mathcal{O}(x)$, which appear at the same PN order as the quadrupole contribution. However, as $x \sim v^2 \ll 1$ for SMBHBs in the PTA band, these terms are always subleading compared to the leading dipole term, in the dipole-dominated regime as well as in the quadrupole-dominated regime.
We will thus neglect them in the following.

The term $Q_{\mathcal{G}}$ represents the leading-order direct contribution of the GB invariant to the EOM. While it appears at $\mathcal{O}(x^5)$, it encodes a further suppression by a factor of $(\lambda/r)^2$. Indeed, scalarization requires black holes to be approximately as large as the coupling scale $\lambda$, implying that $\lambda/r\sim GM/r\sim v^2$. Therefore, we can safely neglect $Q_{\mathcal{G}}$. Nevertheless, the couplings of the scalar field to powers of the GB invariant are indirectly affecting the PN dynamics, as they set the value of the scalar charges. In summary, the formula we derive for scalar radiation does not depend on the specific EsGB interaction $F(\varphi) f(\mathcal{G})$.  Therefore, all the results presented in this paper should be valid for a broad class of theories, as long as the phenomenological fit for the scalar charge in Eq.~\eqref{eq:tildealpha} remains applicable. 

Moreover, we introduced the term
\begin{equation}
\begin{split}
    W=&\frac{(\alpha_A+\alpha_B)\alpha_B'}{(1+q)(1+\alpha_A\alpha_B)}\Delta\alpha\left(\alpha_A-q\frac{\alpha_A'}{\alpha_B'}\alpha_B\right)\,.
\end{split}
\end{equation}
This contribution vanishes if at least one of the two BHs is outside the scalarization window, as, in that case, both the scalar charge and its derivatives vanish. On the other hand, if both the BHs are scalarized, we can use the further approximation
\begin{equation}
    \Delta\alpha\left(\alpha_A-q\frac{\alpha_A'}{\alpha_B'}\alpha_B\right)\simeq\Delta\alpha^2\,,
\end{equation}
as the two BHs, in order to be within the window, must have a comparable size. Hence, this term introduces a subleading contribution $\sim\mathcal{O}(x)$ to the dipole term proportional to $\Delta\alpha^2$, so that we will also neglect it.
All in all, the total radiated power, including both GWs and scalar radiation, is
\begin{equation}
\begin{split}
    \dot E_\mathrm{tot} &\simeq \frac{x^4}{3 G_\mathrm{eff}(1+\alpha_A \alpha_B)} \bigg[ \eta^{2/5} \Delta\alpha^2  \\
    &+\frac{4x}{5} \big(24+(\alpha_A+\alpha_B)^2 \big)+\mathcal{O}\left(x^2,\frac{\lambda}{r}x\right) \bigg]\,.
\end{split}
\label{app:etot}
\end{equation}

Hence, one can compute the general expression for the orbital frequency evolution, by imposing the balance law
\begin{equation}
    \dot{E}_{{\rm tot}}=-\dot{E}_B\,.
\end{equation}
The expression for the GW frequency evolution, which is twice the orbital one, can be finally be obtained as: 
\begin{equation}
\begin{split}
    \frac{{\rm d} f_s}{{\rm d}t} \simeq &\frac{ \,x^{9/2}}{\pi G_{\text{eff}}^2\mathcal{M}^2(1+\alpha_A\alpha_B)}\left(\eta^{2/5}\Delta\alpha^2\right.\\
    &\left.+\frac{4}{5}\left(24+(\alpha_A+\alpha_B)^2\right)x\right)\,,
\end{split}
\end{equation}
We conclude this section with a final remark. Given that we are interested in the early inspiral, we implicitly introduced the further assumption that the scalar charges are constant as the orbit evolves. This appears to be a fairly reasonable approximation in most of the scalarization models, except for the ones that exhibit dynamical scalarization~\cite{Palenzuela:2013hsa,Sampson_2014,Sennett:2016rwa,Julie:2023ncq}. In other words, if the scalar charges are already present at large separations, they generally do not experience a significant enhancement during early-inspiral phase. We do not have any reason to expect that the inclusion of higher powers of the GB invariant significantly affects this assumption.\\

\section{BAYESIAN MODEL COMPARISON} \label{app:bayes}
In the main text, we reported a mild excess in the posterior of $\log_{10}\lambda_9$, suggesting a slight preference for the EsGB theory over GR in explaining PTA data. However, the definitive statistical preference for a model over another is quantified by the Bayes factor, defined as the ratio of the \textit{evidences} $Z$ of the two models. In formulae:
\begin{equation}
    B_\text{10} = \frac{p(\log_{10}\delta t | \mathcal{M}_1)}{p(\log_{10}\delta t |\mathcal{M}_0)} \equiv \frac{Z_1}{Z_0},
    \label{eq:bayes}
\end{equation}
where $\mathcal{M}_1$ and $\mathcal{M}_0$ denote the EsGB and GR theories, respectively. As expressed in Eq.~\eqref{eq:bayes}, the Bayes factor $B_{10}$ quantifies whether it is more probable that the PTA data, collected in $\log_{10}\delta t$, are generated from EsGB ($\mathcal{M}_1$) or from GR ($\mathcal{M}_0$). A value $B_{10} > 1$ favors EsGB, $B_{10} < 1$ favors GR, and $B_{10} = 1$ indicates that the two models are equally likely.

The Bayes factor in Eq.~\eqref{eq:bayes} can be rewritten as
\begin{equation}
    B_{10} = \frac{\int \mathcal{L}(\log_{10}\delta t|\theta_1, \mathcal{M}_1)~ \pi(\theta_1|\mathcal{M}_1)~d\theta_1}{\int \mathcal{L}(\log_{10}\delta t|\theta_0, \mathcal{M}_0)~ \pi(\theta_0|\mathcal{M}_0)~d\theta_0},
    \label{eq:bayesmarg}
\end{equation}
where the evidences are expressed as the product of the \textit{likelihoods} and the \textit{priors}, integrated over the model parameters $\theta_1$ and $\theta_0$. Incidentally, this also highlights why the evidence is also referred to as \textit{marginalized likelihood}. The likelihood for EsGB is given in Eq.~\eqref{eq:lik_EPTAnew}, as a function of the parameters $\theta_1 = \{\log_{10} \frac{\dot{n}_0}{\text{Mpc}^{-3}\text{Gyr}^{-1}}, \log_{10} \lambda_9, \beta, z_0, \gamma, \log_{10}\mathcal{M}_\star/\text{M}_\odot\}$. For GR, the likelihood has the same functional form, but with $\lambda_9 = 0$, resulting in the set of parameters $\theta_0 = \{\log_{10} \frac{\dot{n}_0}{\text{Mpc}^{-3}\text{Gyr}^{-1}}, \beta, z_0, \gamma, \log_{10}\mathcal{M}_\star/\text{M}_\odot\}$. For simplicity, we will consider only the case where the mass ratio $q$ follows a log-uniform distribution over $\log_{10}q\in[-1.5, 0]$;  using a uniform distribution yields similar results.

The integrals in Eq.~\eqref{eq:bayesmarg} are evaluated via Monte-Carlo integration. 
In practice, the evidences are approximated by drawing $N$ random realizations of the collection of parameters $\theta_k$ for the model $k = 0,1$ and computing 
\begin{equation}
    Z_k \sim \frac{1}{N}\sum_{i=1}^{N} \mathcal{L}(\log_{10}\delta t|\theta_{ki}, \mathcal{M}_k),
    \label{eq:zk}
\end{equation}
where $\theta_{ki}$ is the $i$-th random draw of the parameters for the model $\mathcal{M}_k$.
This approximation assumes that the priors on $\theta_k$ for the model $\mathcal{M}_k$ are normalized and uniform within the integration interval, which is verified in our analysis.

\begin{figure}
    \centering
    \includegraphics[width = 0.45\textwidth]{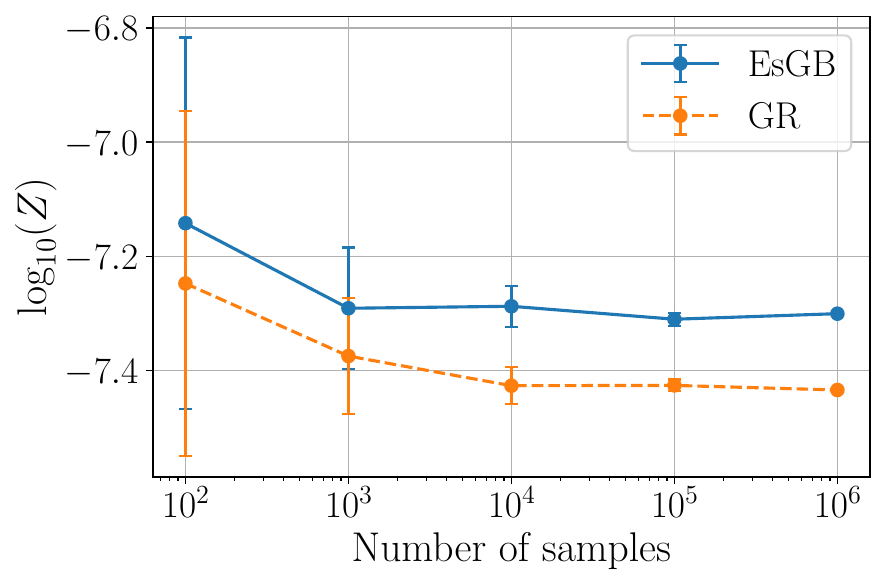}
    \caption{Monte-Carlo estimate of the evidence $Z$ with associated uncertainty $\sigma_Z$, as computed by Eqs.~\eqref{eq:zk} and~\eqref{eq:sigmazk}, for an increasing number of random samples. The solid blue line is computed assuming the EsGB theory of gravity, while the dashed orange line shows the evidence for GR. The plot is produced for the EPTADR2New dataset, using $N_\text{bins} = 9$.}
    \label{fig:montecarlo}
\end{figure}

Given the estimation of $Z_k$ through Eq.~\eqref{eq:zk}, we compute the associated uncertainties as 
\begin{align}
    &\sigma_k^2 = \frac{1}{N}\sum_{i=1}^N\bigg[\mathcal{L}(\log_{10}\delta t|\theta_{ki}, \mathcal{M}_k) - Z_k\bigg]^2\\
    &\sigma_{Z_k}^2 = \frac{1}{N}\sigma^2_k.\label{eq:sigmazk}
\end{align}
Figure~\ref{fig:montecarlo} illustrates this procedure for the EPTADR2New dataset with $N_\text{bins} = 9$. The evidences for EsGB and GR are shown in solid blue and dashed orange, respectively, and are computed using Eqs.~\eqref{eq:zk} and~\eqref{eq:sigmazk} for an increasing number of random samples. Convergence is reached for $N \gtrsim 10^4$, and the error bars follow the scaling predicted by Eq.\eqref{eq:sigmazk}.

Figure~\ref{fig:montecarlo} clearly shows that the EsGB theory explains the data slightly better than GR, as the Bayes factor in Eq.~\eqref{eq:bayes} turns out to be larger than one. However, substituting the converged values in Eq.~\eqref{eq:bayes}, we find $B_{10} \sim 1.35$ for a large number of samples, corresponding to a ``barely worth mentioning'' preference, according to Jeffreys'~\cite{Jeffreys:1939xee} and Kass and Raftery's~\cite{Kass01061995} scales. In other words, even though the Bayes factor mildly favors EsGB over GR, the preference is largely inconclusive. Future data will be essential to support a more definitive conclusion. We performed the same analysis for the NANOGrav 15-year dataset and for the EPTADR2New with $N_\text{bins} = 4$, obtaining similar results. The corresponding plots are omitted for brevity.
\\

To further confirm our findings, we also compute the Bayes factor using the Savage-Dickey density ratio~\cite{Dickey:1971, Verdinelli:1995, joram_soch_2025}. This method is applicable when one model is nested within the other - that is, when the former can be derived by setting certain parameters of the latter to fixed values. This is exactly our case: indeed, GR is obtained by setting $\lambda_9 = 0$ in EsGB theory or, equivalently, $\log_{10}\lambda_9 = -\infty$. Formally, we can write $\theta_0 = \theta_1|_{\log_{10}\lambda_9 = -\infty}$. 

The further requirement $\pi (\theta_0|\mathcal{M}_0) = \pi(\theta_0|\log_{10}\lambda_9 = -\infty,\mathcal{M}_1)$ is trivially satisfied in our case, as the priors are fully separable. 
In this case, the Bayes factor reads:
\begin{equation}
    B_{10} = \frac{\pi(\log_{10}\lambda_9 = -\infty|\mathcal{M}_1)}{\mathcal{P}(\log_{10}\lambda_9 = -\infty|\log_{10}\delta t, \mathcal{M}_1)},
\end{equation}
and is given by the ratio of the prior and posterior probability densities at $\log_{10}\lambda_9 = -\infty$, both evaluated in the EsGB model.

The advantage of this method lies in its efficiency: it avoids computing $Z_0$ and $Z_1$ separately, which can be computationally expensive, and requires only prior and posterior samples from the more general model $\mathcal{M}_1$.

The intuition behind this approach is that if the data favor EsGB over GR, the posterior density at $\log_{10}\lambda_9 = -\infty$ becomes smaller than the prior density at that point,  resulting in a Bayes factor greater than one.\\

In our analysis, the prior on $\log_{10}\lambda_9$ does not extend all the way to $-\infty$; however, values of $\log_{10}\lambda_9$ close to the lower end of the prior range are effectively indistinguishable from GR for our purposes. 

Therefore, as a rough approximation, we evaluate the Savage-Dickey density ratio at $\log_{10}\lambda_9  = -5$. For the EPTADR2New dataset with $N_\text{bins} = 9$, this yields $B_{10} \sim 1.2$, which is broadly in agreement with the result obtained via Monte-Carlo direct integration. Accordingly, in the main text we only report Bayes factors obtained from direct evidence integration, noting that they are consistent with alternative estimation methods.
\\
\section{SUPPLEMENTAL PLOTS} \label{app:other_plots}
\begin{figure}
    \centering
    \begin{subfigure}[t]{0.45\textwidth}
        \centering
        \includegraphics[width=\textwidth]{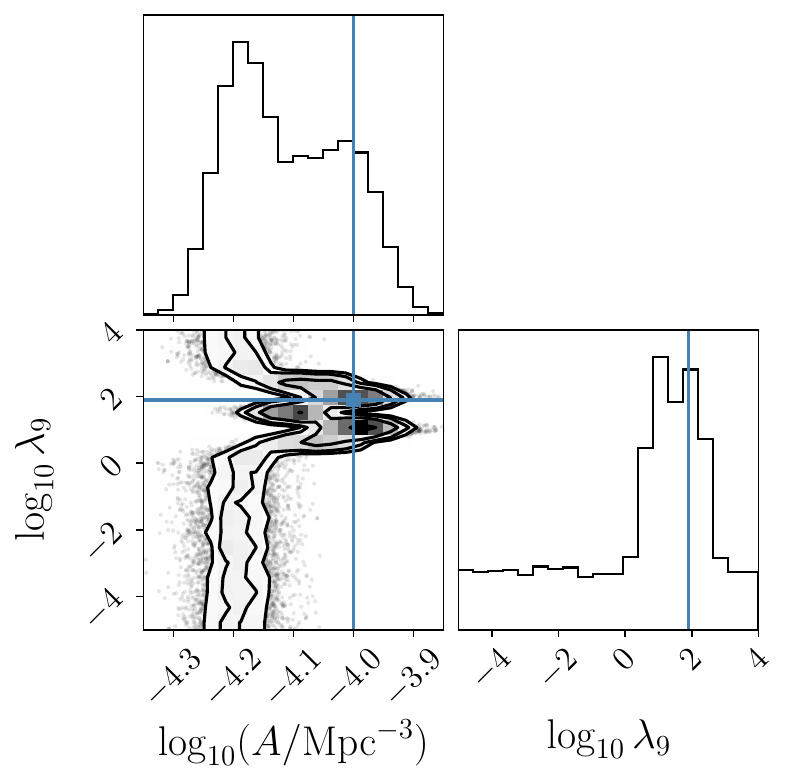}
    \end{subfigure}
    \begin{subfigure}[t]{0.45\textwidth}
        \centering
        \includegraphics[width=\textwidth]{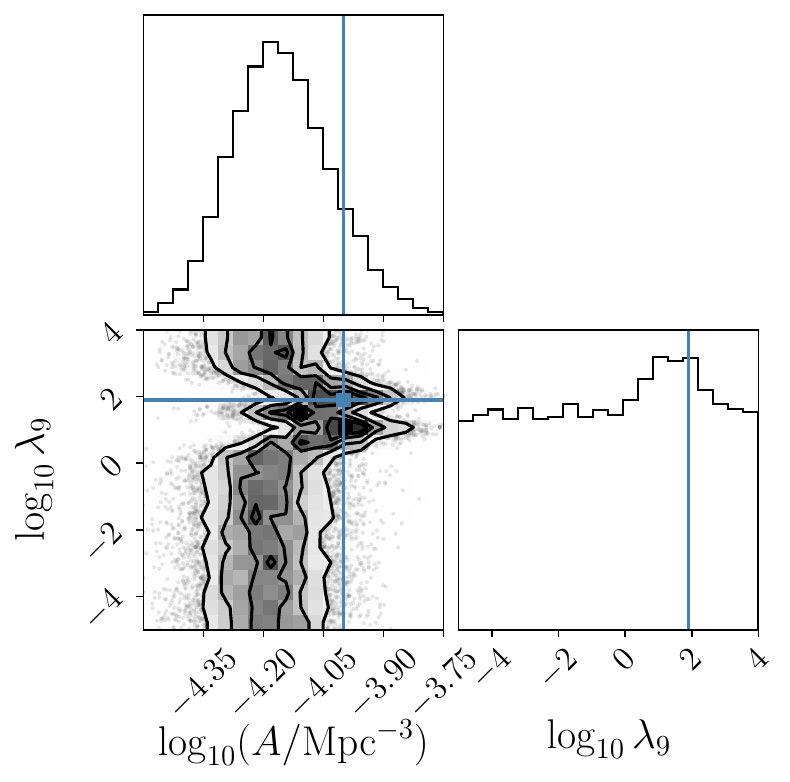}
    \end{subfigure}
    \caption{Posteriors on $\log_{10} (A/\text{Mpc}^{-3})$ and $\log_{10} \lambda_9$ for the idealized analysis described in Sec.~\ref{sec:idealized}. In the upper plot, we set $\epsilon = 1/50$, while the lower plot shows $\epsilon = 1/20$. The blue solid lines show the injected values.}
    \label{fig:app_chains_5020}
\end{figure}

\begin{figure}
    \centering
    \includegraphics[width = \textwidth]{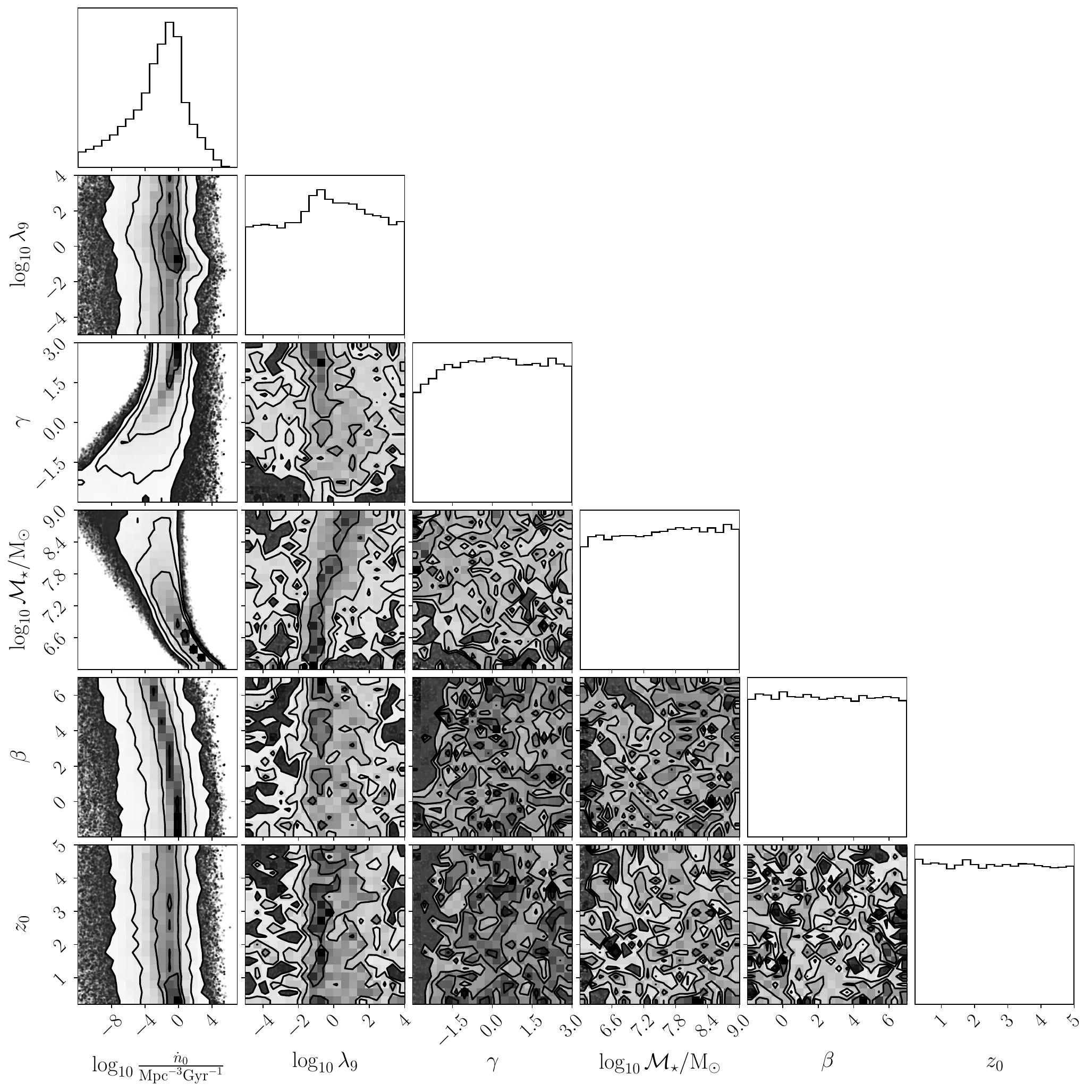}
    \caption{Posteriors on $\{\log_{10} \frac{\dot{n}_0}{\text{Mpc}^{-3}\text{Gyr}^{-1}}, \log_{10} \lambda_9, \beta, z_0, \alpha, \log_{10}\mathcal{M}_\star/\text{M}_\odot\}$ for the analysis of the EPTA DR2New data described in Sec.~\ref{sec:realPTA}, using the first four frequency bins. }
    \label{fig:EPTADR2New_4}
\end{figure}

\begin{figure}
    \centering
    \includegraphics[width = \textwidth]{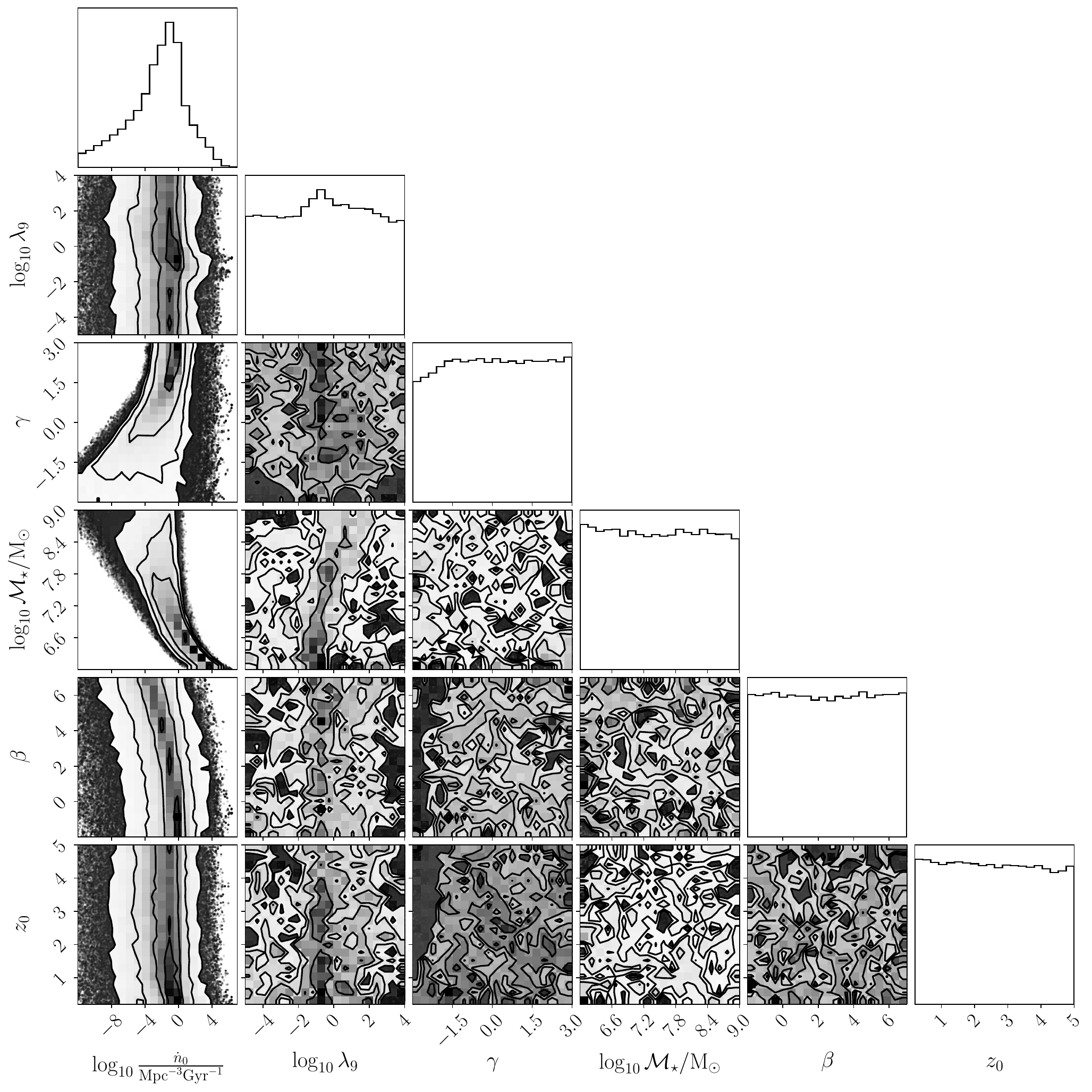}
    \caption{Posteriors on $\{\log_{10} \frac{\dot{n}_0}{\text{Mpc}^{-3}\text{Gyr}^{-1}}, \log_{10} \lambda_9, \gamma,  \log_{10}\mathcal{M}_\star/\text{M}_\odot, \beta, z_0\}$ for the analysis of the NANOGrav 15yr data, as described in Sec.~\ref{sec:realPTA}, using the first fourteen frequency bins. }
    \label{fig:NANOGrav}
\end{figure}

In this section, we present some plots that integrate the discussion in the main text. In Fig.~\ref{fig:app_chains_5020}, we show the posteriors on $\log_{10} (A/\text{Mpc}^{-3})$ and $\log_{10} \lambda_9$ for the idealized analysis described in Sec.~\ref{sec:idealized}. The upper panel shows $\epsilon = 1/50$, the lower panel shows $\epsilon = 1/20$. As $\epsilon$ increase, the constraints become wider, signaling a precision loss on the recovery of the injected parameters (solid blue lines). However, the characteristic shape in the 2D corner plot remains, and can be considered as a distinctive signature of EsGB data. Figure~\ref{fig:EPTADR2New_4} presents the posterior distributions of the parameters describing the merger density model in Eq.\eqref{eq:middl}, obtained from the EPTA DR2New dataset  using the first four frequency bins. Finally, Fig.~\ref{fig:NANOGrav} shows the results for the NANOGrav 15yr dataset, using the first fourteen frequency bins.

%\newpage
\onecolumngrid

\twocolumngrid

%\bibliographystyle{unsrtnat}
%\bibliographystyle{ieeetr}
%\bibliography{Bibliography}
\end{document}